\documentclass[showpacs,amsfonts,amssymb,amsmath]{iopart}
\usepackage{iopams}
\usepackage{cite}
\usepackage{ccaption}
\usepackage{graphicx,epsfig,color,bm,subfig,float}
\usepackage[applemac]{inputenc}

\newcommand{\eqs}[2]{equations~(\ref{#1}--\ref{#2})} 
\newcommand{\figs}[2]{figures~\ref{#1} and \ref{#2}} 

\newcommand{\bea}{\begin{eqnarray}}
\newcommand{\eea}{\end{eqnarray}}
\newcommand\be{\begin{equation}}
\newcommand\ee{\end{equation}}

\newcommand{\lang}{\left\langle}
\newcommand{\rang}{\right\rangle}
\renewcommand{\(}{\left(}
\renewcommand{\)}{\right)}

\newcommand{\p}{\partial}
\newcommand\tn{\tilde n}
\newcommand\tV{\tilde V}
\newcommand\tT{\tilde T}
\newcommand\vth{v_{{\rm th}}}
\newcommand\vthi{v_{{\rm th,i}}}
\newcommand\vthe{v_{{\rm th,e}}}
\newcommand\vtha{v_{{\rm th,a}}}
\newcommand\kpar{k_\parallel}

\newcommand{\uv}{\hat{\bf e}}

\renewcommand{\etal}{\textit{et al.}}

\begin{document}


\title{Electromagnetic effects in the stabilization of turbulence by sheared flow}

\author{M D J Cole$^1$$^,$$^2$ S L Newton$^1$, S C Cowley$^1$, N F Loureiro$^3$, D Dickinson$^1$, C Roach$^1$ and J W Connor$^1$} 

\address{$^1$ EURATOM/CCFE Fusion Association, Culham Science Centre, Abingdon, Oxon, OX14 3DB, UK}
\address{$^2$ Max-Planck-Institut f\"ur Plasmaphysik, EURATOM Association, 17491 Greifswald, Germany}
\address{$^3$ Associa\c{c}\~ao EURATOM/IST, Instituto de Plasmas e Fus\~ao Nuclear -- Laborat\'orio Associado, Instituto Superior T\'ecnico, 1049-001 Lisboa, Portugal}

\ead{michael.cole@ipp.mpg.de}


\begin{abstract}

We have extended our study of the competition between the drive and stabilization of plasma microinstabilities by sheared flow to include electromagnetic effects at low plasma $\beta$ (the ratio of plasma to magnetic pressure).
The extended system of characteristic equations is formulated, for a dissipative fluid model developed from the gyrokinetic equation, using a twisting mode representation in sheared slab geometry and focusing on the ion temperature gradient mode.
Perpendicular flow shear convects perturbations along the field at the speed we denote as $Mc_s$ (where $c_s$ is the sound speed). $M > 1/ \sqrt{\beta}$ is required to make the system characteristics unidirectional and inhibit eigenmode formation, leaving only transitory perturbations in the system. This typically represents a much larger flow shear than in the electrostatic case, which only needs $M>1$.
Numerical investigation of the region $M < 1/\sqrt{\beta}$ shows the driving terms can conflict, as in the electrostatic case, giving low growth rates over a range of parameters. Also, at modest drive strengths and low $\beta$ values typical of experiments, including electromagnetic effects does not significantly alter the growth rates. 
For stronger flow shear and higher $\beta$, geometry characteristic of the spherical tokamak mitigates the effect of an instability of the shear Alfv\'{e}n wave, driven by the parallel flow shear.

\end{abstract}



\section{Introduction}
\label{secintro}

The interplay of sheared flows with fine scale turbulence in tokamaks is the subject of numerous studies, for example~\cite{biglarietal1990,waelbroecketal1992,artunetal1995,diamondetal2005}.
The intense interest stems from the large body of experimental results showing a correlation between flow shear and the suppression of thermal diffusivity -- see for example~\cite{politzeretal2008,fieldetal2011,schaffneretal2012}.
As it has long been known that sheared flows parallel to a magnetic field can be destabilizing~\cite{dangelo1965,cattoetal1973}, this suggests that a clear understanding of the impact of the flow geometry is needed to develop this route to improved tokamak confinement.

The Ion Temperature Gradient (ITG) instability is believed to be a major driver of particle and heat loss from tokamak plasmas~\cite{doyleetal2007}. The effect of flow shear on the stability of the electrostatic ITG mode was investigated previously in a plasma slab with sheared magnetic field.
A simple fluid model retaining collisional dissipation was developed from the gyrokinetic equation~\cite{abeletalxx} and studied using twisting-shearing coordinates~\cite{kelvin,robertstaylor1965,hameirichun1990,waelbroeckchen1991,waltzetal1994,howesetal2001}, which shear simultaneously with the flow in time and the magnetic field in space. Instabilities in this system take the form of twisting eddies. The following important features of the interaction could then be clearly identified~\cite{newtonetal2010}.

As anticipated, parallel flow shear produces an additional driving term for instability, along with the usual density and temperature gradients. This is the parallel velocity gradient (PVG) mode, identified in ~\cite{dangelo1965,cattoetal1973}.
Unstable modes localize where the perpendicular gradients are small, so that the dominant collisional dissipation, due to ion perpendicular viscosity, is minimized.
The effect of perpendicular flow shear is to convect perturbations along the system, at the speed $u_f$ (defined in equation~\ref{equf}). The Mach number $M$ is defined as the ratio of this convection speed to the sound speed.

For $M > 1$, the convection is so strong that the propagation of the system characteristics becomes unidirectional. Conventional unstable eigenmodes can then no longer form. All perturbations are swept along the system, towards regions of strong dissipation resulting from the strong shear.
Such transiently growing perturbations may reach sufficient amplitude to trigger sustained turbulence nonlinearly, as suggested by Waelbroeck~\etal~\cite{waelbroecketal1994}. This regime of subcritical turbulence is well known in the fluid dynamics community~\cite{orr1907,orszagpatera1980,trefethenetal1993,grossmann2000} and its signature appears in gyrokinetic studies of tokamak turbulence~\cite{roachetal2009,highcocketal2010,barnesetal2011,parraetal2011}.

Here we consider the extension of the previous study to include electromagnetic effects, again starting from the gyrokinetic description~\cite{abeletalxx}.
The effects of small but finite $\beta$ are now retained, where $\beta$ is the ratio of the stored plasma thermal energy to the magnetic energy of the externally applied confining magnetic field.
This allows for the propagation of shear Alfv\'{e}n waves through the system, while compressional Alfv\'{e}n waves are still ordered out by the usual gyrokinetic ordering.
From the local dispersion relation, the possibility for an Alfv\'{e}nic instability driven by parallel flow shear may be identified. This is damped here by further retaining the small parallel dissipation due to electron-ion collisions.

Typical values of $\beta$ in tokamaks are low, in the region of 1 - 10\%, so the shear Alfv\'{e}n propagation speed is significantly faster than the sound speed. We find that the Mach number must now be greater than $1/\sqrt{\beta}$ to reach the regime where only transiently growing instabilities are possible, with eigenmodes unable to form. This again corresponds to unidirectional propagation of the characteristics, but will typically represent a substantial increase in the required flow shear compared to the electrostatic limit.

We have therefore investigated the region $0 < M < 1/\sqrt{\beta}$ in detail numerically. 
We recover the electrostatic case in the appropriate limit, and also find that at finite $\beta$, the interplay of the ITG and PVG drives continues to produce regions of weak instability when $M$ is well below the limit of convective stabilization.
For the moderate gradient and $\beta$ values typical of current operating regimes, we find that the inclusion of electromagnetic effects does not significantly alter the behaviour of the growth rates.
However, with increasing $\beta$ and drive strengths, the details of the flow geometry can strongly influence the impact of the new Alfv\'{e}nic instability on the system.

In~\sref{secsyseqns} we introduce the extension of the system equations to allow for electromagnetic effects. We restrict the investigation to the linear regime. Upon neglecting dissipation, the local dispersion relation is obtained in~\sref{secemldr} and the additional Alfv\'{e}nic instability of the system identified. The numerical investigation of the system is presented in~\sref{secnumerics}, demonstrating the behaviour of eigenmodes at $M < 1/\sqrt{\beta}$ and the onset of transient instabilities for $M > 1/\sqrt{\beta}$. We close with a brief summary and discussion in~\sref{secconclusions}.


\section{System equations}
\label{secsyseqns}

We use a dissipative fluid model to study the electromagnetic modification of ITG mode stability in the presence of flow shear.
This is developed from the gyrokinetic equation~\cite{abeletalxx,friemanchen1982,sugamahorton1998}, and the derivation of the ion response, particularly the collisional dissipation, was given in detail in Ref.~\cite{newtonetal2010}.
Therefore, in this section we restate the system geometry for clarity, but simply state the minor modifications to the ion response and refer the reader to the previous work for details.
The electron response was previously taken to be adiabatic, and a more detailed response is required to investigate the electromagnetic case. This is discussed below and in~\ref{appendixa}.


\subsection{Geometry}
\label{secgeometry}

\begin{figure}
\centering
\subfloat []{\includegraphics[width=0.47\textwidth]{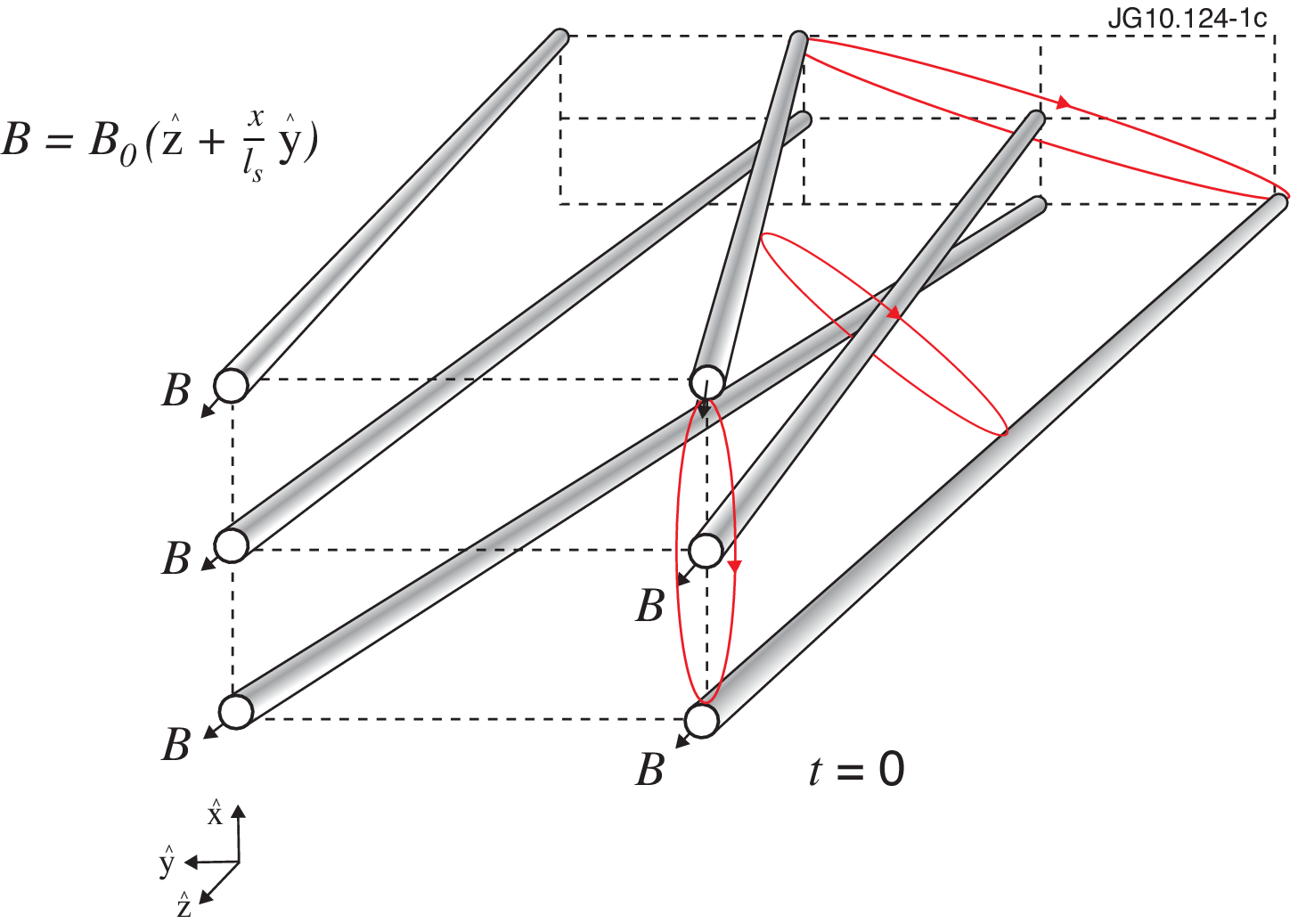}}
\subfloat []{\includegraphics[width=0.40\textwidth]{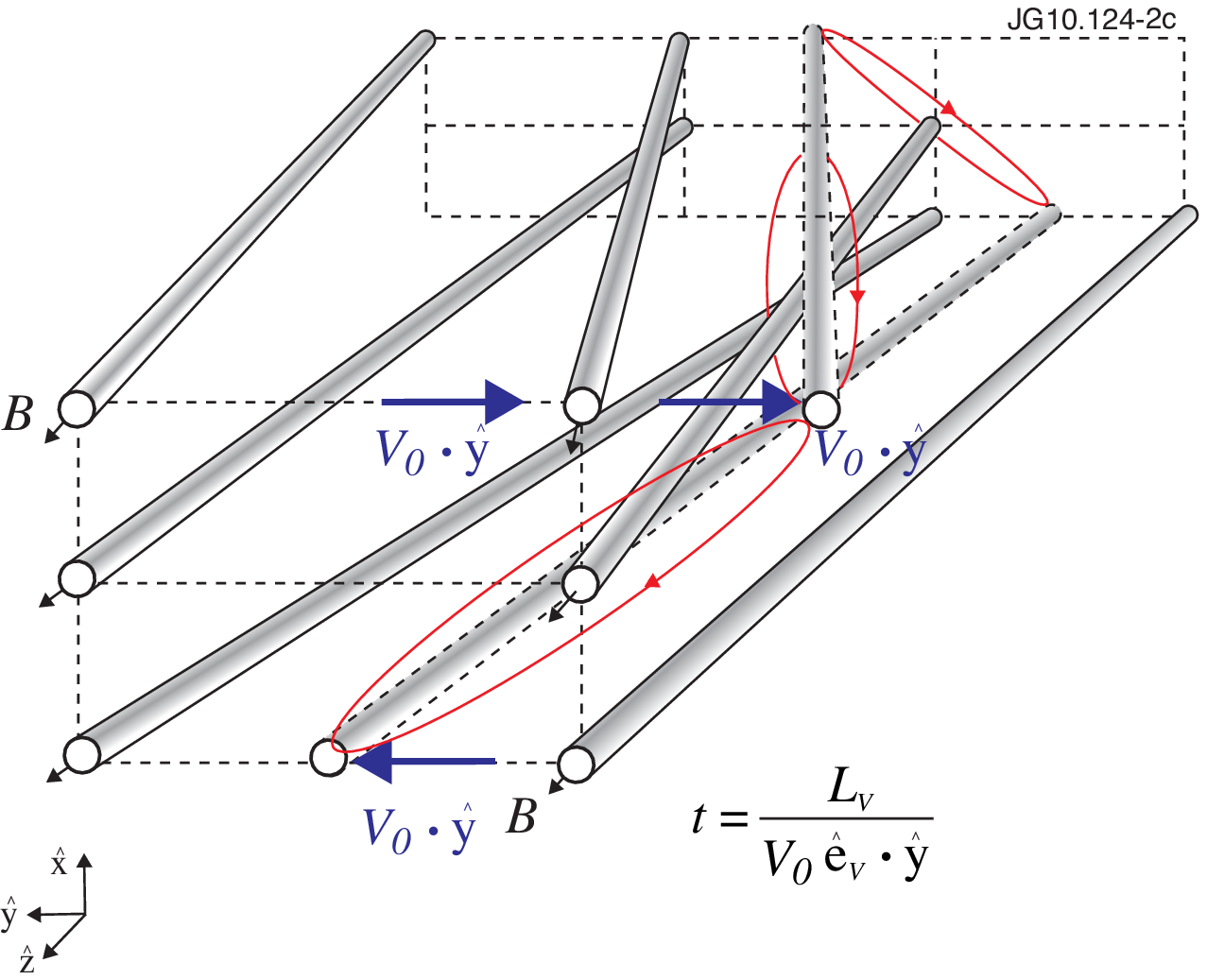}}
\caption{Geometry of the shearing magnetic field ${\bm B}$ and background flow ${\bm V}_0$. Looking in the $-{\bf \hat{z}}$ direction, the field is represented by flux tubes, twisting above and below the plane of $x=0$. Dashed grids are a guide, both $l_s$ and $L_v$ are taken to be negative. (a) Cross sections (red ovals) through a typical eddy at an initial time $t=0$. (b) Flux tubes are advected along $y$ by the perpendicular component of the sheared flow, see \eref{eqcoords}. The eddy is thus twisted and its drive aligned position (eddy parallel to ${\bf \hat{x}}$) retreats along the field at speed $u_f = V_0 \(l_s/L_v\){\bf \hat e_v \cdot \hat y}$. Reproduced from~\cite{newtonetal2010}.}
\label{figuregeom}
\end{figure}

The infinite plasma slab is considered for clarity, with sheared background magnetic field and flows represented in the following forms:
\bea
\bm B &=& B_0\({\bf\hat z} + \frac{x}{l_s}{\bf \hat y}\),\\
\bm V_0 &=& V_0\frac{x}{L_v}\uv_v.
\eea
The magnetic field shears in the ${\hat y}$ direction, and $\uv_v$ is a unit vector lying in the $(y,z)$-plane, in the direction of the flow, as shown in~\fref{figuregeom}. The magnetic field and flow vary in the $x$-direction, with characteristic scale lengths $l_{s}$ and $L_{v}$ respectively, and we have chosen a frame such that there is no flow on the $x=0$ surface.
As in~\cite{newtonetal2010}, we will separate the effects of the parallel and perpendicular components of the velocity shear.
The background density and temperature gradient drive instability as usual, and are also taken to be in the $x$-direction.
As discussed in~\cite{newtonetal2010} the instability is taken to be localized to the region where $\left(x/l_s,x/L_v\right) \ll 1$, so the field and flow profiles studied here can be thought of as Taylor expansions around $x=0$ of more general profiles.

To simplify the equations, we transform to the doubly sheared coordinate system~\cite{robertstaylor1965,howesetal2001,newtonetal2010}
\be
t' = t,	\qquad
z' = z + u_{f}t,	\qquad
y' = y - \frac{x}{l_{s}}z',
\qquad x' = x.
\label{eqcoords}
\ee
With $u_f =0$, this defines the set of field line following coordinates in the presence of a sheared magnetic field and so aligns the coordinate lines with the characteristics of the plasma response.
The perpendicular component of the flow shear introduces a time dependent shear. It convects any structure along the system, with the effective velocity $-u_f$ where
\be
\label{equf}
u_{f} = V_{0}\frac{l_{s}}{L_{v}}\uv_{v}\cdot {\bf \hat y}.
\ee
The coordinate system follows the magnetic field lines as they are twisted by the combined effect of shear in the field lines and background perpendicular flow shear:
\bea
\bm B \cdot \nabla x^{\prime} = \bm B \cdot \nabla y^{\prime} = 0, \qquad
\bm B \cdot \nabla \equiv B_0 \frac{\p }{\p{z^{\prime}}}, \\
\frac{\p }{\p t} + {\bm V_0\cdot{\bf \hat y}}\frac{\p }{\p y} = 
\frac{\p }{\p t^{\prime}} +  u_f\frac{\p }{\p z^{\prime}}. 
\eea

Perturbations in the system take the form of eddies, localized in $x$ and extended along the field, twisting to remain on the surfaces of constant $y^\prime$, along which the plasma has its characteristic response. This is illustrated in~\fref{figuregeom}. This structure is twisted with time by the perpendicular flow shear, so the drive aligned section of the mode, where it lies parallel to $\hat{\bf x}$, retreats along the field with speed $u_f$. When this is sufficiently fast, the plasma response is too slow to form a fully extended mode, and the structure shown becomes transitory.


\subsection{Plasma response}
\label{secplasmaresponse}

The equations used to describe the plasma response will now be outlined.
A simple, quasineutral plasma is considered, with a single hydrogenic ion species of charge $e$ and mass $m_i$. The equilibrium density of both species is $n_{0}$, and the equilibrium ion and electron temperatures are taken to be approximately equal and denoted by $T_0$. We consider electromagnetic instabilities, with perturbation frequency, $\omega$, much less than the ion cyclotron frequency, $\Omega_i = eB_0/m_i$ and phase speed comparable to the ion thermal speed.

The response of both ions and electrons is developed from the gyrokinetic equation which determines the gyroaveraged non-adiabatic piece of the distribution function and is given for a sheared slab including electromagnetic contributions in Ref.~\cite{abeletalxx}. Therefore the usual gyrokinetic ordering~\cite{friemanchen1982,sugamahorton1998} is implicit as a primary expansion
\be
\frac{\omega}{\Omega} \sim \mathcal O\(\frac{k_\parallel}{k_\perp}\) \sim 
\mathcal O\(\frac{\delta f}{F_0}\)\ll 1.
\label{eqgkorder}
\ee
Parallel and perpendicular are taken with respect to the equilibrium magnetic field, $\delta f$ is the total perturbed distribution function, $F_0$ represents the bulk distribution, which is Maxwellian, and $k$ is the wavenumber of the perturbation. Thus compressional Alfv\'{e}n waves are ordered out of the analysis and instabilities can be described by the two potential fields $\phi$ and $A_\parallel$, such that the perturbed electric and magnetic fields are: $\delta {\bf E} = -\nabla \phi - \p_t A_\parallel {\bf b}$ and $\delta \bm B = \nabla \times A_\parallel {\bf b}$, with the unit vector ${\bf b} = {\bm B}/B$.

As in Ref.~\cite{newtonetal2010} we consider the collisional limit for simplicity, so make a subsidiary expansion for both species with (dropping species subscripts for clarity):
\be
\nu~\gg~\omega,~\omega^*,~\vth k_\parallel,~u_f k_\parallel,~\nu k^2 \rho^2. 
\label{eqsuborder}
\ee
Here $\nu$ is the self collision frequency of the species, $\omega^*$ represents the drift frequencies associated with the background gradients (see~\eqs{eqdrives}{eqdriftfreqs}), $\vth=\(2T_0/m\)^{1/2}$ is the species thermal velocity and $\rho = \vth/\Omega$, is the species gyroradius.
Expanding the gyrokinetic equation with respect to $\omega/\nu$, we obtain to lowest order a perturbed Maxwellian distribution (see~\ref{appendixa} and Ref.~\cite{newtonetal2010}).  
The perturbed density, $\delta n$, parallel velocity, $\delta V_\parallel$, and temperature, $\delta T$, of this Maxwellian obey the three fluid conservation equations derived from the density, parallel momentum and energy moments of the gyrokinetic equation.
All quantities on the right hand side of \eref{eqsuborder} are treated as the same order in the derivation of the ion fluid equations, so collisional dissipation is retained.  
Note that $k\rho_i \sim \mathcal O \(\sqrt{\omega/\nu_i}\)\ll 1$. The details are given in Ref.~\cite{newtonetal2010} and lead to~\eqs{eqnpert}{eqtpert} below, modified here only by the inclusion of the terms describing the perturbed magnetic field, represented by the electromagnetic potential $A_\parallel$.

The electron thermal velocity is much larger than the phase speed of the instabilities of interest, so the perturbed parallel electron velocity is zero to leading order. The moment equations describing the density and temperature evolution, via the divergence of the parallel electron flow and thermal flux, may then be replaced by the direct evaluation of these flows from Spitzer-type functions~\cite{spitzerharm1953,hsbook}. The derivation of the electron response is given in~\ref{appendixa}. For simplicity, we retain $T_e \approx T_i = T_0$ and a finite background ion temperature gradient, but neglect the electron temperature gradient locally, so electron temperature perturbations may be neglected, $\delta T_e = 0$. The parallel electron momentum equation is also conveniently replaced by the vorticity equation~\eref{eqapert}, which here relates the ion polarisation current to the divergence of the parallel current, allowing the propagation of shear Alfv\'{e}n waves in the system. The parallel current is 
dissipated by resistivity, which is modelled here by a basic Krook electron-ion collision operator~\cite{cattoetal1979,hahmchen1985}.

As usual the adiabatic piece of the perturbed distribution function (see for example~\ref{appendixa}) must be added to $\delta n$ to give the total density perturbation, $\delta n^t$, of a species. Requiring quasineutrality~\eref{eqphipert} of the perturbed electron and ion densities then gives a closed set of equations for the evolution of the five perturbed variables describing the system: the three fluid variables $\delta n^t$, $\delta V_{\parallel,i}$ and $\delta T_i$, which are associated with the entropy and sound waves in the electrostatic limit, and the two field variables $\phi$ and $A_\parallel$, which here allow the shear Alfv\'{e}n wave. We again use the following normalizations, typical of many gyrokinetic codes (see for example~\cite{GS2}), which emphasize the features of the unstable modes: rapid perpendicular and long parallel spatial dependence, the characteristic acoustic timescale and the amplitude scaling.
\bea
\fl x^{\prime}= \rho_s \tilde x, \qquad
& y^{\prime}=\rho_s \tilde y, \qquad
& z^{\prime}= l_s \tilde z, \qquad
t = \frac{l_s}{c_s} \tilde t, \label{eqnormscales}\\
\fl \tV = \frac{\delta V_{\parallel,i}}{c_s}\frac{l_s}{\rho_s}, \qquad
& \tT = \frac{\delta T_i}{T_0}\frac{l_s}{\rho_s}, \qquad
& \tn = \frac{\delta n^t}{n_0}\frac{l_s}{\rho_s}, \\
\fl \tilde{\phi} = \frac{e \phi}{T_0}\frac{l_s}{\rho_s},\qquad
& \tilde{A}_\parallel = \frac{e c_s A_\parallel}{T_0}\frac{l_s}{\rho_s},
\eea
where $\rho_s = c_s/\Omega$ is the sound Larmor radius associated with the 
sound speed $c_s = \sqrt{\(\gamma_e + \gamma_i\)T_0/m_i}$. 
In this collisional model,  $\gamma_e = 1$ (isothermal) and $\gamma_ i = 5/3$ 
(adiabatic).
We also introduce the Mach number, $M$, associated with the moving frame
\be
M = \frac{u_f}{c_s}.
\ee
The tildes denoting the final transformations will now be dropped 
for convenience.

As in Ref~\cite{newtonetal2010} we restrict ourselves to investigating the linear evolution of the system. All fields are taken to vary as $\exp(i k y)$ multiplied by a function of $z$, so the perpendicular gradient operator becomes: $\nabla_\perp^2 = -k_\perp^2 = - k^2 \( 1 + z^2\)$.
The background gradients provide the instability drive.
The scale lengths for the equilibrium density, parallel velocity and ion temperature are:
\be
\frac{1}{l_n} = \frac{d}{dx}\ln n_0, \qquad
\frac{1}{l_v} = \frac{1}{L_v}\frac{V_0}{c_s} {\bf \hat e_v} \cdot {\bf \hat{z}} \qquad
\frac{1}{l_T} = \frac{d}{dx}\ln T_0,
\label{eqdrives}
\ee
so we use the following effective drift frequencies to characterise the relative strength of the drives:
\be
\omega_n^* = \frac{3}{8}k \frac{l_s}{l_n}, \hspace{1cm} \omega_T^* = \frac{3}{8}k \frac{l_s}{l_T}, \hspace{1cm} \omega_v^* = \frac{3}{8}k \frac{l_s}{l_v}.
\label{eqdriftfreqs}
\ee

The linearized set of five equations describing perturbations of the system therefore takes the final form:
\bea
\label{eqnpert}
\fl & \left(\frac{\p}{\p{t}} + M \frac{\p}{\p{z}}\right)n + \frac{\p}{\p{z}}V = i \omega_n^* \phi - i \omega_v^* A_\parallel,
\\
\label{eqvpert}
\fl & \left(\frac{\p}{\p{t}} + M \frac{\p}{\p{z}}\right)\left(V + \frac{3}{8}A_\parallel\right) + \frac{3}{8}\frac{\p}{\p{z}}\left(n + \phi + T\right) = i \omega_v^* \phi - i \frac{3}{8}\left(\omega_n^* + \omega_T^*\right) A_\parallel \nonumber \\ \fl & \hspace{8cm} - \nu_k \left(1+z^2\right) V,
\\
\label{eqtpert}
\fl & \left(\frac{\p}{\p{t}} + M \frac{\p}{\p{z}}\right)T + \frac{2}{3}\frac{\p}{\p{z}}V = i \omega_T^* \phi - i \frac{2}{3}\omega_v^* A_\parallel - \chi_k \left(1+z^2\right) T,
\\
\label{eqphipert}
\fl & \left(\frac{\p}{\p{t}} + M\frac{\p}{\p{z}}\right)A_\parallel - \frac{\p}{\p{z}}\left(n - \phi\right) = i \omega_n^* A_\parallel - \eta_\parallel \left[\frac{k^2}{\beta}\left(1+z^2\right)A_\parallel - \frac{8}{3} V\right],
\\
\label{eqapert}
\fl & \left(\frac{\p}{\p{t}} + M \frac{\p}{\p{z}}\right) \phi  + \frac{1}{\beta k_\perp^2}\frac{\p}{\p{z}}\left(k_\perp^2 A_\parallel\right) = -  \left(i \omega_n^* + i \omega_T^*\right)\phi + i \omega_v^* A_\parallel \nonumber \\ \fl & \hspace{7cm} - \frac{z}{\left(1+z^2\right)}\left[V + M \left(n + \phi + T\right)\right].
\eea
The electromagnetic parameter is defined here as the square of the ratio of the sound to Alfv\'{e}n speeds
\be
\beta = \frac{c_s^2}{v_A^2} = \frac{8T_0/3m_i}{B^2/\mu_0m_in_0}.
\label{eqbeta}
\ee
The normalized diffusive ion viscosity and thermal conductivity were derived in Ref.~\cite{newtonetal2010}, giving $\nu_k=k^2 \nu_\perp$ and 
$\chi_k=k^2 \chi_\perp$, where
\be
\label{eqnu_chi_perp}
\left(\nu_\perp,\chi_\perp\right) = \left(\frac{9}{40},\frac{1}{4}\right)
\sqrt{\frac{2}{3}}\frac{l_s n_0 e^4 \ln \Lambda}{8 \pi^{3/2} \epsilon_0^2T_0^2 }.
\ee
The parallel resistivity arising from electron-ion collisions is derived in~\ref{appendixa} and takes the analogous normalized form
\be
\label{eqresnormalised}
\eta_\parallel = \frac{\sqrt{3}}{4}\sqrt{\frac{m_e}{m_i}}\frac{l_s n_0 e^4 \ln \Lambda}{8 \pi \epsilon_0^2T_0^2},
\ee
so
\be
\label{eqeta_norm}
\frac{\eta_\parallel}{\nu_\perp} = 6.8 \times 10^{-2}.
\ee
The dissipative terms can be seen to increase strongly for higher $k$ and with distance, $z$, along the field line.
%


\subsection{Characteristics}
\label{seccharacteristics}

The~\eqs{eqnpert}{eqapert} describing the linear system can be written clearly in characteristics form. The characteristics represent the five basic waves present. Defining the following combinations of perturbations:
\bea
S &=& \frac{3}{2}T - n, \qquad
C_\pm = V \pm \frac{3}{4}\left(n + \frac{T}{2}\right), \\
\psi_\pm &=& \phi \pm \frac{1}{\sqrt{\beta}}A_\parallel + \frac{1}{4}S - \frac{1}{2 \left(1 \mp \sqrt{\beta}\right)}C_+ + \frac{1}{2 \left(1 \pm \sqrt{\beta}\right)} C_-,
\eea
we obtain:
\bea
& \left[\frac{\p}{\p{t}} + M\frac{\p}{\p{z}}\right]S  = S^d, \label{eqschar} \\
& \left[\frac{\p}{\p{t}} + \left(M \pm 1\right) \frac{\p}{\p{z}}\right] C_\pm = C_\pm^d, \label{eqcchar} \\
& \left[\frac{\p}{\p{t}} + \left(M \pm \frac{1}{\sqrt{\beta}}\right) \frac{\p}{\p{z}}\right] \psi_\pm = \psi_\pm^d, \label{eqpsichar}
\eea
with the drives for the evolution of the amplitudes given by:
\bea
\fl S^d = \frac{i}{2}\left(\frac{3}{2}\omega_T^* - \omega_n^*\right) \left(\psi_+ + \psi_- + \frac{1}{\left(1-\beta\right)}\left(C_+- C_-\right) - \frac{S}{2}\right) \nonumber \\
\fl \hspace{5cm}  - \frac{1}{2}\chi_k\left(1 + z^2\right) \left(C_+ - C_- + \frac{3}{2}S\right), \\
\fl C_\pm^d = \frac{i}{2} \left(\omega_V^* \pm\frac{3}{4}\left(\omega_n^* +\frac{\omega_T^*}{2}\right)\right) \left(\psi_+ + \psi_- + \frac{1}{\left(1 - \beta\right)}\left(C_+ - C_-\right) - \frac{S}{2}\right) \nonumber \\
\fl \hspace{1cm} \mp \sqrt{\beta}\frac{i}{2} \left(\omega_V^* \pm\frac{3}{4}\left(\omega_n^* +\frac{\omega_T^*}{2}\right)\right) \left(\psi_+ - \psi_- + \frac{\sqrt{\beta}}{1 - \beta}\left(C_+ + C_-\right)\right) \nonumber \\
\fl \hspace{1cm} - \frac{1}{2}\nu_k\left(1 + z^2\right)\left(C_+ + C_-\right) \mp \frac{1}{8}\chi_k\left(1+z^2\right)\left(\frac{3}{2}S + C_+ - C_-\right) \nonumber \\
\fl \hspace{1cm} + \frac{3}{8} \eta_\parallel \left[ \frac{k^2}{2 \sqrt{\beta}}\left(1 + z^2\right)\left[\psi_+ - \psi_- + \frac{\sqrt{\beta}}{{1 - \beta}}\left(C_+ + C_-\right)\right] - \frac{4}{3}\left(C_+ + C_-\right)\right], \\
\fl \psi_\pm^d = \nonumber \\ \fl -\frac{i}{4}\left[\frac{5}{2}\omega_n^* + \frac{5}{4}\omega_T^* + \frac{1}{\left(1-\beta\right)}\left(\pm2\sqrt{\beta}\omega_V^* + \frac{3}{2}\left(\frac{\omega_T^*}{2} + \omega_n^*\right)\right)\right] \nonumber \\
\fl \hspace{5cm} \times \left(\psi_+ + \psi_- + \frac{1}{\left(1 - \beta\right)}\left(C_+ - C_-\right) - \frac{S}{2}\right) \nonumber \\
\fl \pm \sqrt{\beta}\frac{i}{2}\left[\frac{\omega_n^*}{\sqrt{\beta}} + \frac{3}{4}\frac{\sqrt{\beta}}{\left(1-\beta\right)}\left(\omega_n^* +\frac{\omega_T^*}{2}\right) \pm \frac{2-\beta}{\left(1-\beta\right)}\omega_{V}^*\right]\left(\psi_+ - \psi_- + \frac{\sqrt{\beta}}{\left(1-\beta\right)}\left(C_+ + C_-\right)\right) \nonumber \\
\fl - \frac{z}{\left(1 + z^2\right)}\left[\frac{1}{2}\left(C_+ + C_-\right) + \frac{1}{\sqrt{\beta}}\left(\psi_+ - \psi_- + \frac{\sqrt{\beta}}{\left(1-\beta\right)}\left(C_+ + C_-\right)\right) \right. \nonumber \\ \left. + M \left[\frac{1}{2}\left(\psi_+ + \psi_-\right) + \left(\frac{5}{6} + \frac{1}{2\left(1-\beta\right)}\right)\left(C_+ - C_-\right)\right]\right] \nonumber \\
\fl + \frac{1}{8} \frac{\beta}{\left(1-\beta\right)}\chi_k\left(1+z^2\right)\left(C_+ - C_- + \frac{3}{2}S\right) \pm \frac{1}{2}\frac{\sqrt{\beta}}{\left(1-\beta\right)}\nu_k\left(1+z^2\right)\left(C_+ + C_-\right) \nonumber \\ 
\fl \mp \frac{\eta_\parallel}{\sqrt{\beta}} \left( 1 + \frac{3}{8}\frac{\beta}{1-\beta}\right) \nonumber \\
\fl \hspace{2cm} \times \left[\frac{k^2}{2 \sqrt{\beta}}\left(1 + z^2\right)\left[\psi_+ - \psi_- + \frac{\sqrt{\beta}}{{1 - \beta}}\left(C_+ + C_-\right)\right] - \frac{4}{3}\left(C_+ + C_-\right)\right].
\eea
The first three characteristics are the entropy mode propagating the perturbed specific entropy, $S$, at speed $M$, the forward propagating sound wave propagating $C_+$ at speed $M+1$, and the backward propagating sound wave propagating $C_-$ at speed $M-1$. These waves are also present in the electrostatic case considered previously, but the evolution of their amplitude along the characteristic is modified here by the effects of finite $\beta$. In addition, there are now two more waves allowed in the electromagnetic system, the forward propagating Alfv\'{e}n wave, propagating $\psi_+$ at speed $M+1/\sqrt{\beta}$, and the backward propagating Alfv\'{e}n wave propagating, $\psi_-$ at speed $M - 1/\sqrt{\beta}$. The terms on the right hand sides couple the five waves but do not change their propagation speeds.

We may now make the same argument as in the electrostatic case~\cite{newtonetal2010}: when the flow shear is sufficient that the Mach number $M$ is larger than the fastest characteristic speed of the system, all structures will be swept along the system and therefore no unstable eigenmodes may form. This can be seen as follows. Any initial perturbation which is localized in $z$ between $z = a$ and $z = b$ at the time $t = 0$ (i.e. the function describing the initial perturbation has compact support in $z$) must then be localized between $z = a + (M - 1/\sqrt{\beta})t_1$ and $z = b + (M + 1/\sqrt{\beta})t_1$ at time $t = t_1$. Eigenmodes can form when $M < 1/\sqrt{\beta}$, by the combination of oppositely travelling waves, but if $M$ increases, we can see that the speed of the forward travelling characteristics are increased, whilst those of the backward travelling characteristics are reduced. When $M > 1/\sqrt{\beta}$ is reached, all characteristics of the system move forward, due to the convective effect of 
the perpendicular shear of the background flow. An initial perturbation is swept forward, since $z = a + (M - 1/\sqrt{\beta})t_1$ and $z = b + (M + 1/\sqrt{\beta})t_1$ both increase with the time $t_1$, and no eigenmode can form in the system. This behaviour is indeed observed in the numerical investigation of the system, presented in~\sref{secnumerics}, (see~\fref{transients}). Any unstable perturbation will therefore be swept along the system into the dissipative region, where it will be forced to decay. Such a perturbation would be able to grow exponentially for a finite time, and its behaviour would not be captured by a traditional eigenvalue analysis. As noted in the introduction, this could provide a subcritical route to turbulence.

However, to reach this transient limit, at the typical low $\beta$ values relevant to tokamak devices, requires a significant increase in the effective perpendicular flow shear as compared to the electrostatic case. With only the entropy and sound waves present, no eigenmode could be formed for $M > 1$. We can therefore see that electromagnetic effects result in a new range of $M$ values, $1 < M < 1/\sqrt{\beta}$, where an eigenmode may form and at low $\beta$ values , the new region of potential instability is large. However, the couplings on the right hand side are proportional to $\beta$ and therefore will be weak in this case, so we may expect that new, strong instabilities will not arise. The behaviour in this region is investigated in detail numerically in~\sref{secnumerics}, where the characteristic equations are simulated for wide ranges of the drive strength and $\beta$, to map out the impact of the electromagnetic effects in the various ranges of $M$.


\section{Local dispersion relation}
\label{secemldr}

By neglecting the dissipative terms and dropping any explicit dependence on $z$, we can effectively analyze the region close to $z=0$ and establish the nature of the basic instabilities present in the system.
In this section we therefore take such reduced versions of ~\eqs{eqnpert}{eqapert} and look for plane wave solutions of the form $\exp(-i\omega t+i\kpar z)$. The simple convective effect of the perpendicular flow shear, $M$, may be removed by defining a modified frequency in the laboratory frame
\be
\omega^\prime = \omega - k_\parallel M.
\ee
The local dispersion relation is then
\bea
\label{eqemldr}
\fl \left[\omega^{\prime 2} \left(\omega^\prime + \omega_n^*\right) - \omega k_\parallel \left(k_\parallel - \omega_v^*\right) + \frac{k_\parallel^2}{4}\left(\frac{3}{2}\omega_T^* - \omega_n^*\right)\right] \nonumber \\
\fl \hspace{4cm} \times \left[ \omega^\prime\left(\omega^\prime - \omega_n^* - \omega_T^* \right)- k_\parallel\left(\frac{k_\parallel}{\beta} - \omega_v^* \right)\right] = 0.
\eea
In the limit $\beta \rightarrow 0$, this reduces to simply the first factor on the left equal to zero, which is the dispersion relation of the electrostatic coupled ITG-PVG instability, given previously in Ref.~\cite{newtonetal2010}.
Finite $\beta$ introduces the Alfv\'{e}n wave, which has the basic dispersion relation $\omega^{\prime 2} - k_\parallel^2 / \beta =0$ here.
The effect of finite $\beta$ on the ITG mode has been much studied for $M=\omega_v^*=0$. The decoupled local dispersion relation~\eref{eqemldr}, in the limit $k_\perp \rho_i \ll 1$ used here, may be found, for example, in~\cite{kimetal1993,reynders1994,snyder1999,hintonetal2003}. Note that the structure of the ITG component varies slightly due to the different parallel closures employed.
The local stability limits of the ITG-PVG system remain here as determined previously~\cite{newtonetal2010}. The Alfv\'{e}n component is given by the second factor in~\eref{eqemldr} equal to zero. For $\omega_v^* = 0$ this has the solution~\cite{mikhailovskii1972}:
\be
\omega^\prime = \pm \frac{k_\parallel}{\sqrt{\beta}}\sqrt{1 + \frac{\beta \left(\omega_n^* + \omega_T^*\right)^2}{4 k_\parallel^2}} + \frac{\omega_n^* + \omega_T^*}{2},
\ee
so finite $\beta$ does not destabilize the wave in the presence of only density and temperature gradients.

However, in the limit $\omega_n^*=\omega_T^*=0$, the local dispersion relation~\eref{eqemldr} reduces to
\be
\omega^\prime\left[\omega^{\prime2} - k_\parallel\left(k_\parallel - \omega_v^*\right)\right]\left[\omega^2 - k_\parallel \left(\frac{k_\parallel}{\beta} - \omega_v^*\right)\right] = 0.
\label{eqwvldr}
\ee
The first bracketed term is the usual quadratic dispersion relation for the electrostatic PVG instability~\cite{cattoetal1973,newtonetal2010}, which gives instability when
\be
\frac{\omega_v^*}{k_\parallel} > 1,  
\label{eqpvgstability}
\ee
or equivalently, upon removing the normalisations
\be
\frac{V_0}{c_s} > \frac{8}{3}\frac{k_\parallel}{k_y}\frac{L_V}{\rho_s}.
\ee
We see that finite $\beta$ introduces a new mechanism for instability, as the background parallel velocity gradient can also destabilize the Alfv\'{e}n wave when
\be
\frac{\omega_v^*}{k_\parallel} > \frac{1}{\beta},
\label{eqalfvenstability}
\ee
or equivalently
\be
\frac{V_0}{v_A} > \frac{1}{\sqrt{\beta}}\frac{8}{3}\frac{k_\parallel}{k_y}\frac{L_V}{\rho_s}.
\ee
For typical low $\beta$ values, this limit is significantly higher than that posed by PVG stability, but its effects can be seen in the scan over $\beta$ values performed numerically in~\sref{secnumerics}.
The corresponding kinetic derivation of the dispersion relation~\eref{eqwvldr}, for a collisionless system retaining only $\omega_v^*$, is given in~\ref{appendixb}.
The stability limits of both the fluid and kinetic derivations are in close agreement, noting, as above, the effect of different parallel closures.

The parallel and perpendicular flow shear are related geometrically. As in~\cite{newtonetal2010} we define
\be
\omega_v^* = \alpha M, \qquad
\alpha = \frac{3}{8}k{\bf \frac{\hat{e}_v\cdot \hat{z}}{\hat{e}_v\cdot \hat{y}}},
\label{defnalpha}
\ee
and the angles $\theta$ and $\theta_v$, which give the direction of mode propagation and of the background flow with respect to the magnetic field:
\be
\tan \theta = \frac{k}{k_\parallel}, \qquad \tan \theta_v = {\bf \frac{\hat{e}_v\cdot \hat{y}}{\hat{e}_v\cdot \hat{z}}}.
\ee
Given the onset of convective stabilization for $M > 1/\sqrt{\beta}$, in the case with both $\omega_v^*$ and $M$ positive we obtain the simple bounds on the flow shear which will produce the Alfv\'{e}nic instability
\be
\frac{1}{\sqrt{\beta}} > M > \frac{1}{\beta}\frac{8}{3}\frac{\tan \theta_v}{\tan \theta}.
\ee

The dynamics of this instability can be understood from the linear~\eqs{eqnpert}{eqapert}. We take $M=0$ for clarity, and retain only the $\omega_v^*$ drive, neglecting dissipation and explicit $z$ dependence, as was done to obtain~\eref{eqwvldr}. Consider first the quasineutrality relation~\eref{eqphipert}, removing the normalizations temporarily and taking it in the form given in~\eref{eqqnappendix}:
\be
\frac{\p}{\p{z}}\left(\frac{\delta n^t}{n_0}\right) - \frac{e}{T_0}\left(\frac{\p{\phi}}{\p{z}} + \frac{\p{A_\parallel}}{\p{t}}\right) = 0.
\ee
This represents the deviation from the ideal limit $E_\parallel = -\p_z \phi - \p_t A_\parallel = 0$ due to parallel pressure perturbations. To describe the Alfv\'{e}nic instability, such sound wave effects can be neglected, so we take the further limit $c_s \ll v_A$. The quasineutrality and vorticity equations then decouple from the other equations, taking the (normalized) forms
\bea
\frac{\p}{\p{t}}A_\parallel + \frac{\p}{\p{z}}\phi = 0, \\
\frac{\p}{\p{t}}\phi + \frac{1}{\beta}\frac{\p}{\p{z}}A_\parallel = i \omega_v^* A_\parallel.
\label{eqwvvorticity}
\eea
Together these give the Alfv\'{e}nic component of the dispersion relation~\eref{eqwvldr}. From the vorticity equation, we see that a potential perturbation $\phi$ drives an ion polarization current, due to the finite ion mass. This causes an ion density perturbation, of too high order to appear directly in the ion continuity equation~\eref{eqnpert}.
To maintain charge neutrality, an electron current $j_\parallel$ flows, with associated magnetic potential $\mu_0 j_\parallel = -\nabla_\perp^2 A_\parallel$.
In a uniform plasma, this would sustain a shear Alfv\'{e}n wave.
There is an associated radial perturbation of the magnetic field $\delta B_x = i k A_\parallel$, which causes a further ion density perturbation here, as ions follow the magnetic line with a parallel velocity which varies radially, due to the background flow shear. This is the final term on the right of~\eref{eqnpert}.
Whilst the electrons freely follow the perturbed field, the finite thermal ion Larmor radius prevents the ions doing so exactly, producing a net radial current -- the final term in the reduced vorticity equation~\eref{eqwvvorticity}.
When $k_\parallel$, $k$ and $l_s/l_v$ are all positive, the additional ion density perturbation resulting from the flow shear cannot be compensated by the electron current and instability arises.

It is of interest to consider an alternative limit of~\eqs{eqnpert}{eqapert}, again with only the $\omega_v^*$ drive and $M=0$ for clarity, but for large $z$, which corresponds to conditions far along the field line. 
The terms in $\nu_k$ and $\chi_k$ dominate~\eref{eqvpert} and~\eref{eqtpert}, representing strong ion dissipation due to the sheared system, so the fields $V$ and $T$ tend to zero. Neglecting resistivity, plane wave solutions of the form $\exp(-i\omega t+i\kpar z)$ can again be assumed, giving the dispersion relation
\be
\omega^2 - k_\parallel\left(\frac{k_\parallel}{\beta} - 2 \omega_v^*\right) = 0.
\ee
Thus the Alfv\'{e}nic instability is only weakly coupled to the damped thermal component of the system and survives, in modified form, in the sheared field. Only with the inclusion of finite resistivity in~\eref{eqphipert} does it also decay far along the field.


\section{Numerical Results}
\label{secnumerics}

\begin{figure}
\centering
\includegraphics[width=1.0\textwidth]{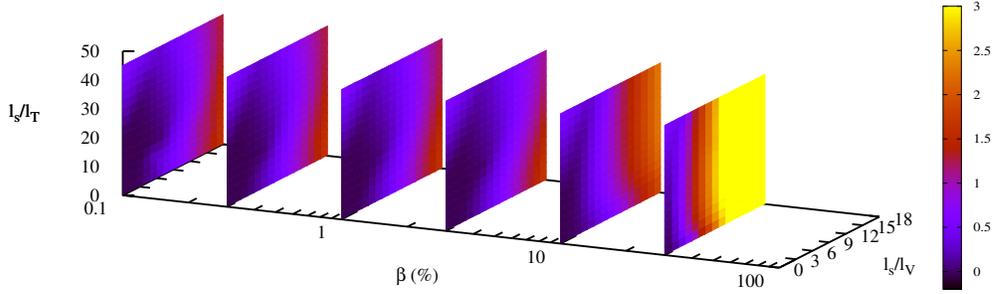}
\caption{Contour plot of the maximum growth rate as a function of the parallel velocity and temperature gradient drives in the range of $\beta=0.001 - 0.3$, taking $M=0.3$ and $l_s/l_n =0$.}
\label{slices}
\end{figure}

In this section, we summarize the results of numerical investigation of the linearized system of~\eqs{eqschar}{eqpsichar}. The characteristic equations were integrated using a second order accurate upwind scheme in a box of size $-z_{inf} < z < z_{inf}$, where $z_{inf}$ was varied between $50$ and $200$ as required to follow the fields until they dissipated. The spatial resolution was $\Delta z = 0.1$, while the temporal resolution was varied to ensure numerical stability, increasing with higher wavenumber, stronger dissipation and decreasing $\beta$.

Convergence tests were performed to verify that the choice of domain size and resolution were sufficient. The perpendicular viscosity is fixed to $\nu_{\perp} = 3.0$ for all cases here, the large value ensuring that the orderings are consistent. The perpendicular diffusivity is then given by~\eref{eqnu_chi_perp} and the parallel resistivity by~\eref{eqeta_norm}. Both the effective drive strengths~\eref{eqdriftfreqs} and the growth rate depend upon the wave number. Therefore the stability of the system is considered here as a function of the physical drives, the normalized gradient scale lengths $(l_s/l_T, l_s/l_v, l_s/l_n)$. Unless stated the growth rate of the fastest growing mode is given, determined using the well-known Brent's method~\cite{NumericalRecipes}. For clarity all the results shown neglect the density gradient drive, $l_s/l_n=0$; at moderate values its inclusion did not qualitatively affect our conclusions.

The numerical results were benchmarked at low $\beta$ against those of the electrostatic case~\cite{newtonetal2010}. Whilst the $\beta \rightarrow 0$ limit of~\eqs{eqschar}{eqpsichar} cannot be simulated directly as the propagation speeds of the Alfv\'{e}nic characteristics become infinite, good agreement is seen already for $\beta \sim 0.001$. This is illustrated in~\fref{slices}, which shows the maximum growth rate at this value of $\beta$ as a function of the temperature gradient, $l_s/l_T$, and parallel flow shear, $l_s/l_v$, drives, for fixed $M = 0.3$. Note this is well below the limit $M_A \approx 32$ for convective stabilization, and eigenmodes with well-defined growth rates form. This may be compared to figure 4 in~\cite{newtonetal2010} -- note that we have confirmed that the region of exactly zero growth shown there at high drive strengths was initially found due to the limitations of the Brent method.
Subsequent slices in~\fref{slices} show the effect on the growth rate of increasing $\beta$. Note that the growth rates are formally evaluated in this section for $\beta$ up to $\sim 1$, but we remember that the limit $\beta \ll 1$ was assumed in the derivation of the system of equations, to eliminate the compressional Alfv\'{e}n wave. As discussed in~\cite{newtonetal2010}, stable or weakly unstable regions occur for a range of combinations of drive strengths even though we remain below the critical threshold in $M$.
The growth rates remain qualitatively similar up to moderate values of $\beta$. However, as $\beta$ is increased above $10 \%$ the area of high growth rate associated with the PVG is strengthened and spreads into the region of higher $l_s/l_T$ where it previously was suppressed by the competing ITG. The strong association of enhanced instability at finite $\beta$ with the parallel flow shear drive suggests a connection with the Alfv\'{e}nic instability described in~\sref{secemldr}.

\begin{figure}
\centering
\includegraphics[width=0.6\textwidth]{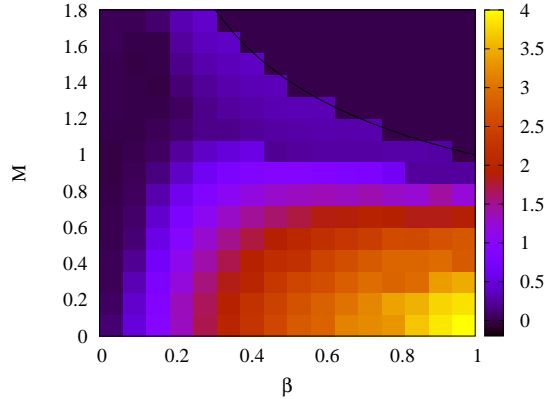}
\caption{Contour plot of the maximum growth rate as a function of $M$ and $\beta$, for $l_s/l_v=5.0$, $l_s/l_T=20.0$, $l_s/l_n=0$. The black line indicates $M = M_A$. Eigenmodes cannot form above this, so the linear growth rate is marked there as zero. Note that subcritical turbulence may still arise in this region.}
\label{mvbfixed}
\end{figure}

As discussed in~\sref{seccharacteristics}, at finite $\beta$ unstable eigenmodes may form when the perpendicular flow shear, characterized by $M=u_f/c_s$ (see~\eref{equf}), is in the range $1 < M < M_A$, where $M_A = 1/\sqrt{\beta}$. In the electrostatic limit, all perturbations must be transient in this region. As $\beta \rightarrow 0$, this range of $M$ increases, but the Alfv\'{e}nic fields $\psi_\pm$ propagate ever more rapidly to large $z$ and dissipate, so the electrostatic limit is recovered, with the growth rates in this region falling to zero. This is demonstrated in~\fref{mvbfixed}, which shows the growth rate as a function of $M$ and $\beta$, at fixed drive strengths. At $\beta = 1$, the electromagnetic, $M=M_A$, and electrostatic $M=1$, convective stability thresholds converge. We see that growth rates become significant in the region $\beta \gtrsim 0.1$, which is consistent with~\fref{slices}.

\begin{figure}
\centering
\subfloat[]{\includegraphics[width=0.4\textwidth]{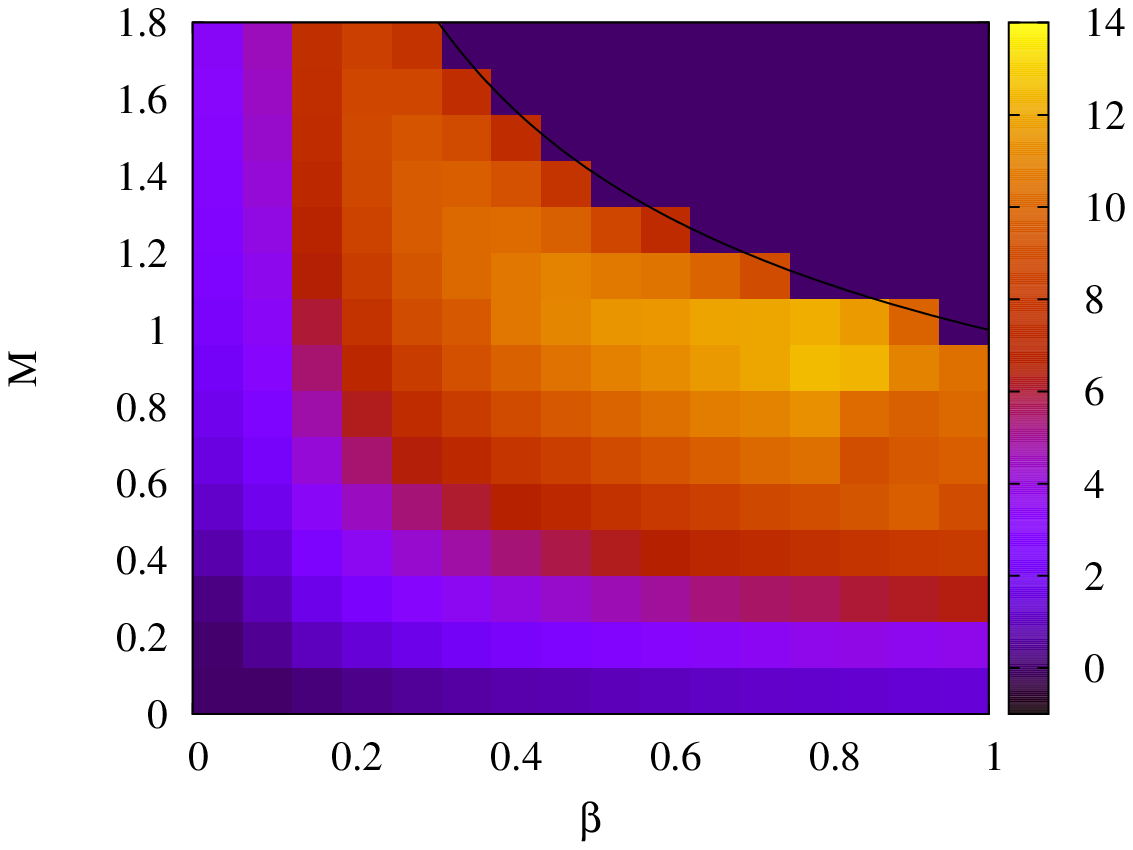}}
\hspace{1cm}
\subfloat[]{\includegraphics[width=0.4\textwidth]{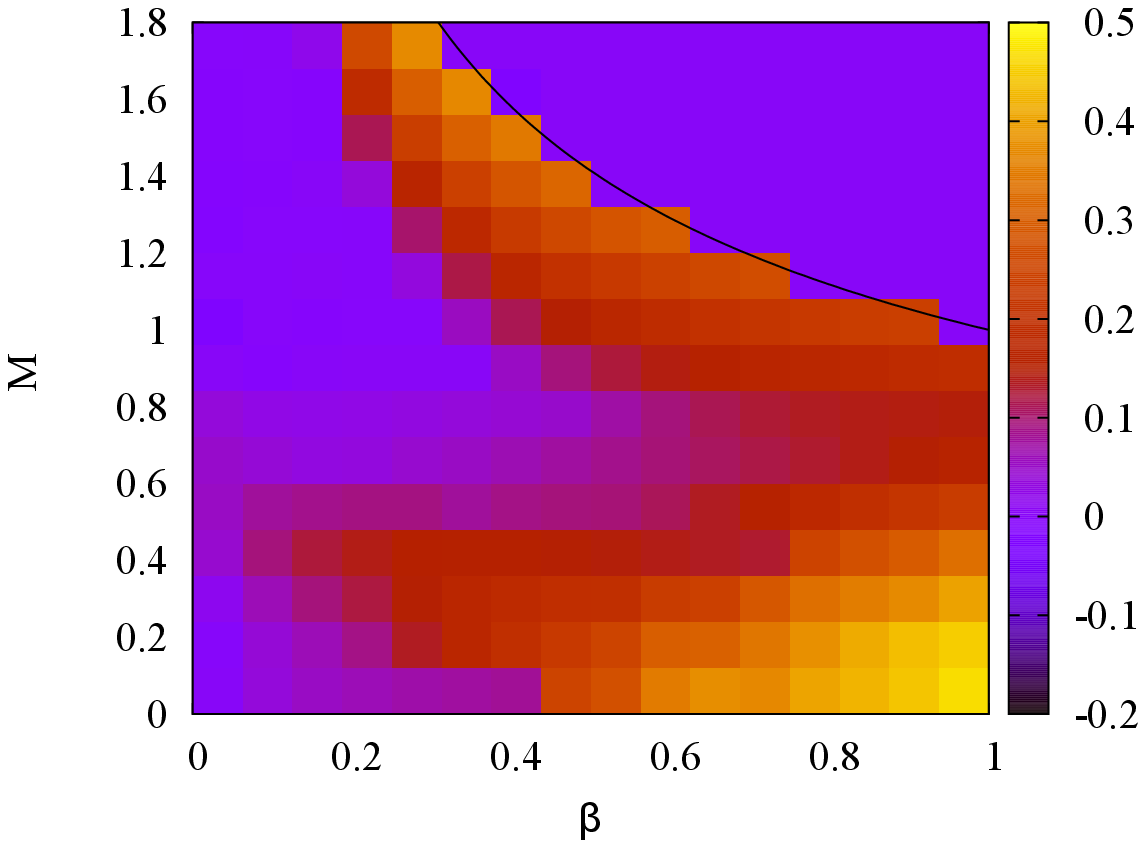}}
\caption{Contour plots of the maximum growth rate as a function of $M$ and $\beta$, for $l_s/l_T=20.0$ and (a) $l_s/l_v=30M$, corresponding to $\theta_v = 2^{\circ}$; (b) $l_s/l_v=M$, corresponding to $\theta_v = 45^{\circ}$. As in \fref{mvbfixed} the black line indicates $M = M_A$, no eigenmodes can form above this and the linear growth rate is therefore marked as zero. Note the different scales of figures (a) and (b).}
\label{mvbflow}
\end{figure}

However, as noted in~\sref{secemldr}, $l_s/l_v$ and $M$ are related geometrically, $l_s/l_v = M \cot \theta_v$. So in~\fref{mvbflow} we consider the more physically realistic case of a varying background flow speed, at fixed angle, $\theta_v$, to the magnetic field. For small $\theta_v$, ~\fref{mvbflow}(a), the background flow is nearly parallel to the background magnetic field, which is typical of a conventional aspect ratio tokamak. At low $M$ the associated PVG drive is weakest and, consistent with~\fref{slices} at this value of $l_s/l_T$, the system is least unstable here across the range of $\beta$ values. Convective stabilization becomes more effective as $M$ increases, but at low $\theta_v$ the effective PVG drive increases much more rapidly, generating strong growth in the wide region of eigenmodes allowed at low $\beta$ (note the different scale to~\fref{mvbfixed}).
Larger $\theta_v$ is more characteristic of tight aspect ratio spherical tokamaks and we illustrate the extreme case of $\theta_v = 45^{\circ}$ in~\fref{mvbflow}(b). Convective stabilization dominates the behaviour by comparison to~\fref{mvbflow}(a), giving more slowly growing instabilities, located in regions similar to those of~\fref{mvbfixed}, the case of fixed PVG drive. Above $M=1$, the system is stable or only weakly unstable for $\beta$ values below $20 \%$.

\begin{figure}
\centering
\subfloat[]{\includegraphics[width=0.4\textwidth]{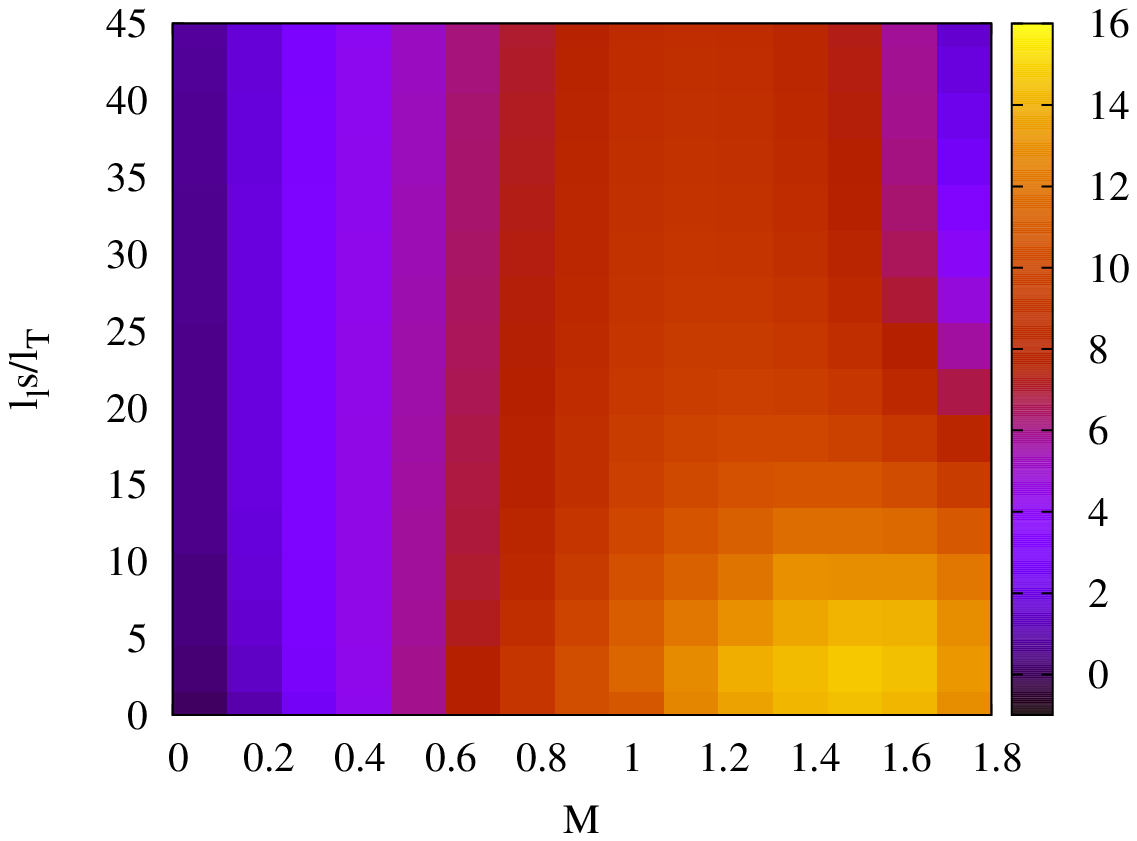}}
\subfloat[]{\includegraphics[width=0.4\textwidth]{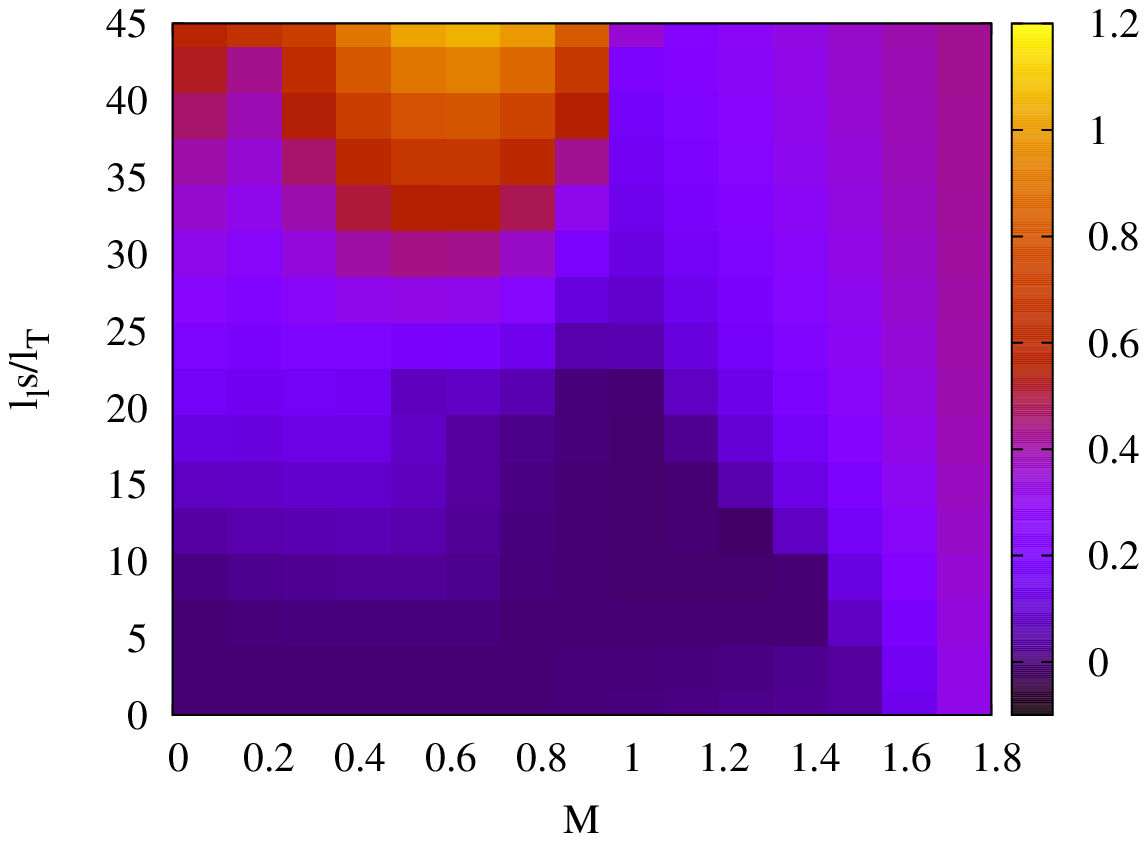}}
\caption{Contour plots of the maximum growth rate as a function of $l_s/l_T$ and $M$, for $\beta=0.3$ and (a) $l_s/l_v=30M$, corresponding to $\theta_v = 2^{\circ}$; (b) $l_s/l_v=M$, corresponding to $\theta_v = 45^{\circ}$. Here $M_A \approx 1.8$ and eigenmodes cannot form at higher $M$ values.}
\label{flowang}
\end{figure}

\begin{figure}
\centering
\subfloat []{\includegraphics[width=0.40\textwidth]{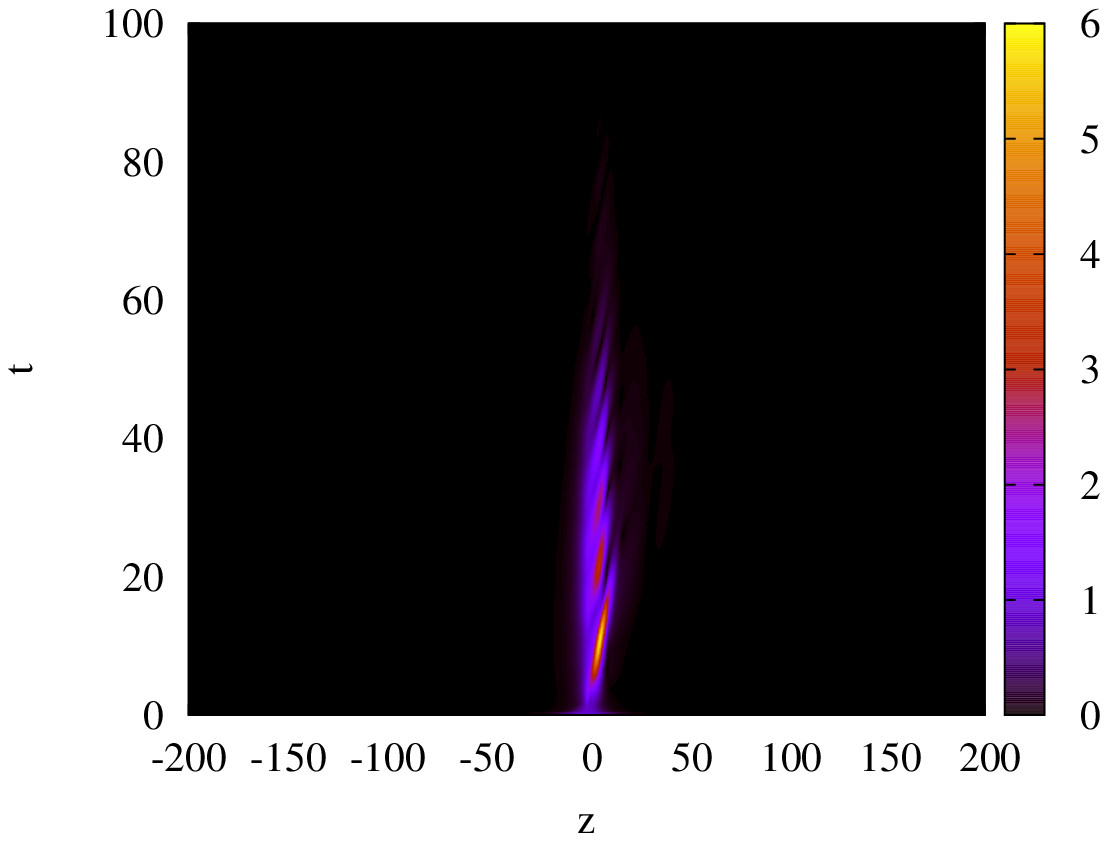}}
\subfloat []{\includegraphics[width=0.40\textwidth]{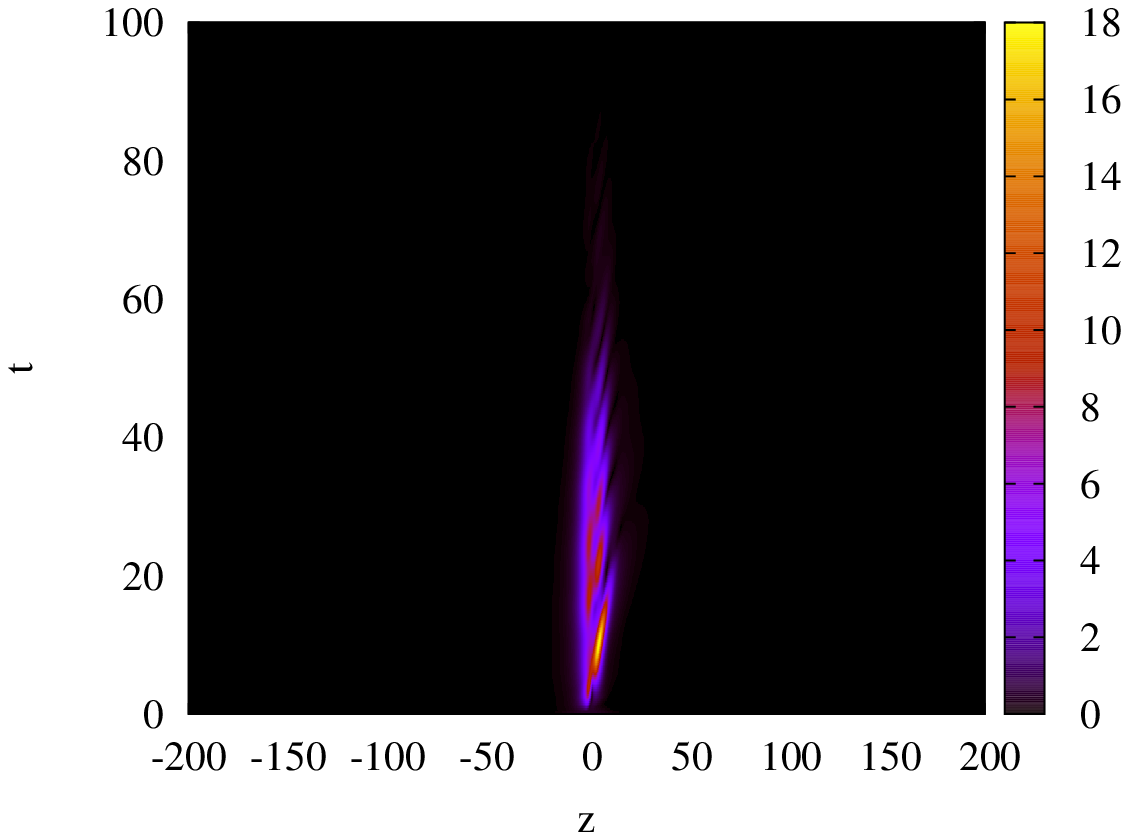}}\\
\subfloat []{\includegraphics[width=0.40\textwidth]{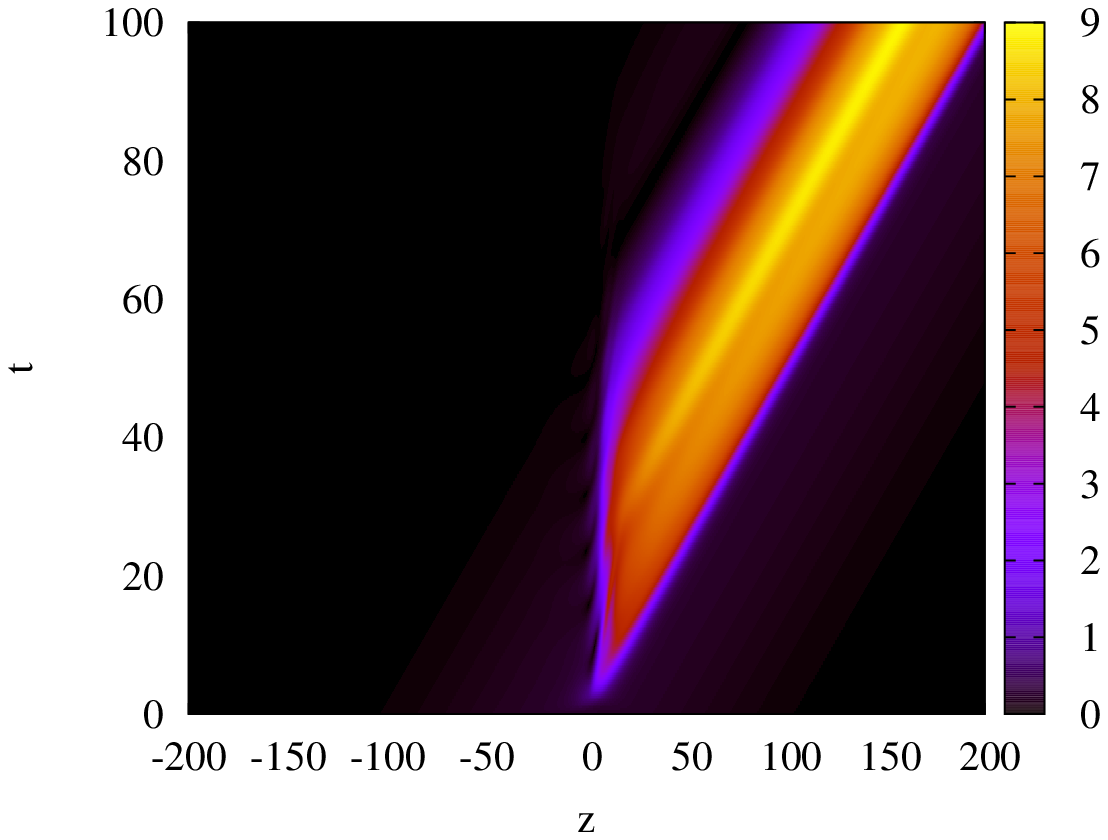}}
\subfloat []{\includegraphics[width=0.40\textwidth]{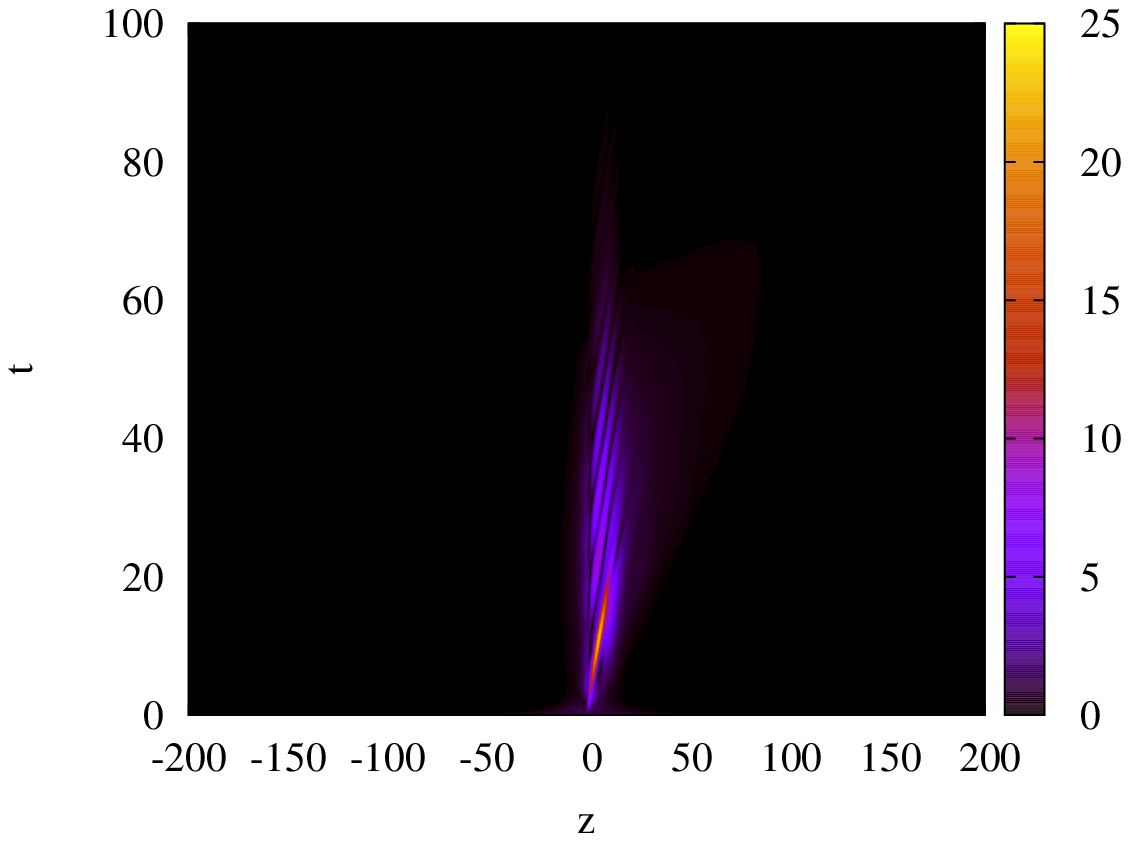}}
\caption{Time history plots of parallel velocity (a), temperature (b), density (c) and parallel magnetic potential (d) perturbations (absolute values) as a function of time, for $M > M_A$, $l_s/l_v = 5.0$, $l_s/l_T=20.0$, $l_s/l_n=0$ and $k=0.1$. Each field is normalized to its maximum amplitude at $t=0$, so colours show the amplification factor of the initial perturbation.}
\label{transients}
\end{figure}

In~\fref{flowang}, we again plot the growth rate with varying flow speed at fixed flow angle, but now for varying ITG drive strength, at fixed $\beta = 0.3$. This may be compared directly with the electrostatic case, figure 5 in~\cite{newtonetal2010}. 
As expected instability now persists above $M=1$ to the threshold at $M = M_A \approx 1.8$ here.
For $\theta_v = 2^{\circ}$, typical of large aspect ratio, inclusion of finite $\beta$ extends the strongly growing region at large $M$, and thus strong PVG drive, to larger values of ion temperature gradient, consistently with~\fref{slices}. The growth rate is also significantly increased at all values of $M$, possibly due to the new Alfv\'{e}nic instability.
For $\theta_v = 45^{\circ}$ the effect of $\beta$ is much less pronounced. As in the electrostatic case, convection acts to produce low growth rates at all but the highest temperature or parallel velocity gradients.
Finally, in~\fref{transients}, we verify that for $M > M_A$ only transient temperature, velocity and parallel magnetic field perturbations exist, which are swept to high $z$ and dissipated. As expected, the density perturbation saturates due to the lack of explicit dissipation in~\eref{eqnpert}. With such strong parallel flow shear, the Alfv\'{e}nic instability described in~\sref{secemldr} was seen to sustain a strongly growing $A_\parallel$ field when resistivity was not included.
%


\section{Discussion and conclusions}
\label{secconclusions}

We have shown previously~\cite{newtonetal2010} that by using appropriate shearing coordinates, basic effects of flow shear on the electrostatic Ion Temperature Gradient (ITG) mode in a sheared slab could be cleanly identified.
With a dissipative fluid model developed from the gyrokinetic equations, it was seen that instabilities in the system took the form of twisting eddies, driven by both the ITG and the parallel component of the velocity gradient (PVG). The perpendicular component of the flow shear convects perturbations along the system at the speed $u_f = Mc_s$, where $c_s$ is the sound speed and $M$ is the Mach number of the moving perturbations~\eref{equf}. This tilts the eddies until collisional dissipation acts to wipe out the strong perpendicular gradients. 
With sufficient flow shear to give $M>1$, this convection makes the propagation of the system characteristics (sound and entropy waves) unidirectional. Eigenmodes can then no longer form and turbulence in such a system would be governed by the behaviour of subcritical fluctuations, which is investigated in~\cite{newtonetal2010,schekochihinetal2012}. Fluctuations with sufficiently large growth rates can lead to a subcritical state with strong turbulence.

Here we have extended the study to include the electromagnetic effects generated by low, but finite, $\beta$. This introduces the shear Alfv\'{e}n wave characteristic, which propagates $1/\sqrt{\beta}$ times faster than the sound wave. 
Correspondingly the critical Mach number to formally eliminate eigenmodes increases from one to $M_A = 1/\sqrt{\beta}$. For typical experimental $\beta$ values of a few percent, this represents a significant increase in the required perpendicular flow shear, compared to the electrostatic case.
A fundamentally electromagnetic instability is also now present in the system, corresponding to destabilization of the shear Alfv\'{e}n wave by the PVG. However, the local analysis indicates that the PVG drive required~\eref{eqalfvenstability} is a factor $1/\beta$ larger than that for the PVG drift mode~\eref{eqpvgstability} and so it should not be a dominant effect in typical operational regimes.

We have investigated numerically the region $0 < M < M_A$, in which eigenmodes can form with well-defined growth rates. Formal results were presented up to $\beta \sim 1$, however we note that the applicability of the system investigated to a plasma must be restricted to the region $\beta \ll 1$.
At low $\beta$, the growth rates fall to zero in the new region of allowed instability $1 < M < M_A$, where previously only transient perturbations were possible and so the electrostatic limit is recovered.
The ITG and PVG drives still compete significantly up to $\beta$ values around 10\%, allowing regions of low growth to be accessed below the convective stability limit. At higher $\beta$, the electromagnetic instability driven by the PVG progressively reduces the region of stability at high drive.
For moderate values of the drive strengths and low $\beta$ values typical of current experiments, the results indicate that including electromagnetic effects does not significantly alter the conclusions drawn from the electrostatic analysis. Precise values for the growth rates however would require a more detailed kinetic treatment.
Finally, it is interesting to consider the behaviour over the wide parameter space in~\figs{mvbflow}{flowang}, which was investigated for two different assumed flow angles, typical of conventional and spherical tokamaks respectively. Whilst a difference in the stability of the two geometries could be identified previously (see~\cite{newtonetal2010}, figure 5), here the more spherical geometry appears robustly more stable, as any applied flow shear produces a strong, stabilizing, convective effect. The electromagnetic instability is driven readily in conventional geometry, where large values of $M$ can only be achieved at the expense of a strong PVG drive.


\ack

The authors wish to thank Matthew Lilley for stimulating discussions 
and Fulvio Militello for assistance in producing Figure 2.
This work was supported in part by the RCUK Energy Programme [grant number EP/I501045], by Funda\c{c}\~ao para a Ci\^encia e 
Tecnologia and by the European Communities under the contracts of Association between EURATOM 
and CCFE and EURATOM and IST.
Simulations were performed at Newhydra (Oxford) and HPC-FF (Juelich).
To obtain further information on the data and models underlying this paper please contact PublicationsManager@ccfe.ac.uk.
The views and opinions expressed herein do not necessarily reflect those of the European Commission.


\appendix


\section{Derivation of electron response}
\label{appendixa}

In this appendix, the derivation of the electron response is detailed. This leads to the equation for the density perturbation~\eref{eqnpert} and the vorticity equation~\eref{eqapert}. 

As in the derivation of the ion response~\cite{newtonetal2010}, we start from the gyrokinetic equation given in Ref.~\cite{abeletalxx}. The species subscript $e$ is dropped here for clarity. We restrict formally to low $\beta$, so the magnetic perturbation is represented by only the parallel component of the magnetic potential, $A_\parallel {\bf b}$ (see~\sref{secplasmaresponse}). The electron distribution function, $f$, correct to first order in the gyrokinetic expansion in $\omega / \Omega$ is
\bea
f = F_0\(\epsilon,{\bm R}\) + \delta f\({\bm r},w_\parallel,w_\perp,\zeta,t\), \\
\delta f = \frac{e \phi\({\bm r},t\)}{T_0} F_0 + h\({\bm R},w_\parallel,w_\perp,t\),
\eea
where $\delta f$ is the perturbation from equilibrium, the background Maxwellian is $F_0 = \(n_0/\pi^{3/2}\vth^3\)\exp\(-\epsilon / T_0\)$, the particle energy is $\epsilon = mw^2 / 2 + e \phi\({\bm r},t\)$, the particle velocity is $\bm v$, the velocity variable ${\bm w} = {\bm v} - {\bm V}_0$ and the guiding center position satisfies ${\bm R} = {\bm r} - {\bf b} \times {\bm v}/\Omega \equiv {\bm r} - \bm \rho$. Electron gyroradius corrections will be neglected, so the distinction between particle and guiding centre position can be dropped. The complications of the finite Larmor radius corrections to the collision operator which were treated in Ref.~\cite{newtonetal2010} therefore do not arise here. The distribution function of gyrocenters, $h$, is independent of the gyroangle of the particle motion, $\zeta$, and is defined by
\bea
\fl \frac{\p h}{\p t} + \( w_\parallel {\bf b} + {\bm V}_0\) \cdot \nabla h + \frac{1}{B_0}\left\{\lang \phi - w_\parallel A_\parallel \rang, h \right\} - \lang C^l\(h\)\rang \nonumber \\
\fl \hspace{2cm} = \frac{F_0}{B_0}\left[\frac{1}{l_n} + \(\frac{\epsilon}{T_0}- \frac{3}{2}\)\frac{1}{l_T} + \frac{2 c_s w_\parallel}{\vth^2 l_v}\right] \frac{\p }{\p y}\lang \phi - w_\parallel A_\parallel \rang \nonumber \\
\fl \hspace{3cm} - \frac{e F_0}{T_0} \left(\frac{\p }{\p t} + {\bm V}_0 \cdot \nabla \right) \lang \phi - w_\parallel A_\parallel \rang . 
\label{hieqEdmund}
\eea
The drive terms appearing here were given in~\sref{secplasmaresponse}, the angled brackets denote the average of the enclosed quantity over the gyroangle: $\lang A \( {\bm r}\)\rang = \(2\pi\)^{-1}\oint A \({\bm R} + {\bm \rho}\) d\zeta$ and the Poisson bracket is defined as: $ \left\{\lang \phi \rang, h \right\} =  \( \nabla \lang \phi \rang \times \nabla h \)\cdot {\bf b}$, where the spatial gradient is taken at constant ${\bm w}$. The linearized electron collision operator appearing here, $C^l$, will be discussed further below. First, the doubly-sheared coordinate transformation outlined in~\sref{secsyseqns} and detailed in Ref.~\cite{newtonetal2010} is implemented. Again we take $k_\perp \rho \ll 1$, so retain, in both the ion and electron responses, only the leading effects of the background gradients - the twist of the eddy structures through $l_s$, the convective motion through $L_V$ (see definition~\eref{equf} for $u_f$) and the parameter gradients $l_n,l_T,l_v$ which drive the drift waves. The 
transformed equation for $h$ then takes the form
\bea
\fl \frac{\p h}{\p t} + \( w_\parallel + u_f\) \frac{\p h}{\p z^\prime} + \frac{1}{B_0}\left\{ \phi - w_\parallel A_\parallel , h \right\} -  C^l\(h\) \nonumber \\
\fl \hspace{2cm} = \frac{F_0}{B_0}\left[\frac{1}{l_n} + \(\frac{\epsilon}{T_0}- \frac{3}{2}\)\frac{1}{l_T} + \frac{2 c_s w_\parallel}{\vth^2 l_v}\right] \frac{\p }{\p y^\prime}\left( \phi - w_\parallel A_\parallel \right) \nonumber \\
\fl \hspace{3cm} - \frac{e F_0}{T_0} \left(\frac{\p }{\p t} + u_f \frac{\p }{\p z^\prime}  \right)\left( \phi - w_\parallel A_\parallel \right).
\label{eqh}
\eea
The Poisson bracket is now taken to represent the transformed value and the primes on the transformed variables are dropped from hereon.

As discussed in~\sref{secplasmaresponse}, we work in the collisional limit. Therefore the distribution function is expanded in $\omega/\nu$ such that $h = h^{(0)} + h^{(1)} + \ldots$, where the order of expansion is denoted by a superscript and the solution for $h$ determined from~\eref{eqh} order by order. The electron self-collision contribution, $C_{ee}^l$, dominates the linearized collision operator, giving to lowest order:
\be
C_{ee}^l\(h^{(0)}\) = 0.
\ee
This indicates~\cite{hmbook,newtonetal2010} that $h^{(0)}$ has the form of a perturbed Maxwellian, with the perturbed density, $\delta n$, temperature, $\delta T$ and parallel velocity, $\delta V_\parallel$
\be
h^{(0)} = \left[\frac{\delta n}{n_0} + \frac{\delta T}{T_0}\(\frac{\epsilon}{T_0} - \frac{3}{2}\) + \frac{2 w_\parallel \delta V_\parallel}{\vth^2} \right] F_0\(w, {\bm R}\). 
\label{eqh0}
\ee
Due to their small mass, the term involving the perturbed parallel velocity is formally negligible for electrons, unlike ions, and will be dropped. The perturbed electron parallel flow is contained in the $h^{(1)}$ piece of the distribution function.

The subsidiary ordering introduced in~\sref{secplasmaresponse} is used to derive the ion response, but must be modified for the electrons. The gyroradius corrections are neglected, as noted, and the phase speed of the instabilities of interest (see~\sref{secplasmaresponse}) is comparable to the ion thermal speed, so $\omega \ll k_\parallel v_{th,e}$. Also, in deriving the ion response, $\phi$ was taken to be comparable to $v_{th,i} A_\parallel$, giving $v_{th,e} A_\parallel \sim \phi \sqrt{m_i/m_e}$ here. Therefore, neglecting non-linear terms hereon, the expansion of~\eref{eqh} to next order is
\bea
\fl \( w_\parallel + u_f\) \frac{\p h^{(0)}}{\p z} - C^{l}\left(h^{(1)}\right) \nonumber \\
\fl = - \frac{F_0}{B_0}\left[\frac{1}{l_n} + \(\frac{\epsilon}{T_0}- \frac{3}{2}\)\frac{1}{l_T}\right] w_\parallel \frac{\p A_\parallel}{\p y} + \frac{eF_0}{T_0} \(\frac{\p}{\p t} + u_f \frac{\p}{\p z} \) w_\parallel A_\parallel,
\label{eqh1exp}
\eea
where all quantities are evaluated at the guiding centre position. Proceeding as in previous work~\cite{newtonetal2010}, we would take moments of this equation and so express the electron density, parallel momentum and energy conservation. However, as there is no flow in $h^{(0)}$, we see that the density and temperature evolution would not be determined at this order, unlike the ion case. As we are considering the collisional limit,~\eref{eqh1exp} may instead be solved directly for $h^{(1)}$ in terms of Spitzer functions, and the two moment equations replaced by direct evaluation of the parallel flow and heat flux, as follows.

Taking the perturbed fields to vary as $\exp\left(i k y\right)$ (see~\sref{secplasmaresponse}) and writing $\delta n$ in terms of the total perturbed density, $\delta n^t$, which will be used in the expression for quasineutrality
\be
\frac{\delta n}{n_0} = \frac{\delta n^t}{n_0} - \frac{e\phi}{T_0},
\ee
the equation for $h^{(1)}$ is
\bea
\fl C^{l}\left(h^{(1)}\right) &&= \left(w_\parallel + u_f\right) \left[\frac{\p}{\p{z}}\left(\frac{\delta n^t}{n_0} - \frac{e \phi}{T_0}\right) + \left(\frac{\epsilon}{T_0} - \frac{3}{2}\right)\frac{\p}{\p{z}}\left(\frac{\delta T_e}{T_0}\right)\right]F_0 \nonumber \\
\fl && \quad + w_\parallel \left[ ik\frac{T_0}{eB_0}\left[\frac{1}{l_n} + \(\frac{\epsilon}{T_0}- \frac{3}{2}\)\frac{1}{l_T}\right]  - \(\frac{\p}{\p t} + u_f \frac{\p}{\p z} \) \right]\frac{eF_0}{T_0} A_\parallel \nonumber \\
\fl && \equiv u_f F_0 \left[A_1 + \(\frac{\epsilon}{T_0}- \frac{3}{2}\)A_2 \right] + w_\parallel F_0 \left[A_3 + \(\frac{\epsilon}{T_0}- \frac{3}{2}\) A_4 \right],
\label{eqh1}
\eea
where the drive terms defined in the final equality are:
\bea
\fl &A_1 = \frac{\p}{\p{z}}\left(\frac{\delta n^t}{n_0} - \frac{e \phi}{T_0}\right),
& \qquad A_3 = A_1 + \frac{e}{T_0} \( i\frac{T_0}{eB_0}\frac{k}{l_n} - \frac{\p}{\p t} - u_f \frac{\p}{\p z} \) A_\parallel, \\
\fl &A_2 = \frac{\p}{\p{z}}\left(\frac{\delta T_e}{T_0}\right),    
& \qquad A_4 = A_2 + i\frac{k}{B_0 l_T} A_\parallel.
\eea
As the equation is linear and $C^{l}\left(F_0\right)=0$, the solution $h^{(1)}$ will be of the form
\be
h^{(1)} = \left(h_1 A_1 + h_2 A_2 + h_3 A_3 + h_4 A_4\right) F_0.
\ee
Frame speeds comparable to the instability phase speeds are of interest, so $h_1,h_2$ will be of order $u_f / v_{th,e}$ smaller than $h_3,h_4$ and we restrict to evaluating only $h_3$ and $h_4$.

As discussed in~\sref{secemldr}, the parallel resistivity introduced by electron collisions is required to damp the parallel current perturbation associated with the Alfv\'{e}n waves in the system and it is sufficient to take a simple electron-ion Krook collision operator for $C^{l}$ in~\eref{eqh1}. More accurate forms of the collision operator could be used, leading to more accurate values for the growth rate in the regions investigated, but at the expense of added complexity and reduced clarity, so they are not pursued here. Introducing
\be
C^{l}\left(h_n F_0 \right) \approx -\nu_{ei} h_n F_0,
\ee
for $n = 3,4$, with  $\nu_{ei} = n_0 e^4 \ln \Lambda / 8 \sqrt{2} \pi \epsilon_0^2 m_e^{1/2} T_0^{3/2}$, gives
\be
\left(h_3,h_4\right) = -\frac{w_\parallel}{\nu_{ei}}\left(1, \frac{\epsilon}{T_0}- \frac{3}{2} \right).
\ee
The parallel flow and heat flux are then
\bea
\fl \delta V_{\parallel,e} & = \frac{1}{n_0} \int w_\parallel \left(h_3A_3 + h_4A_4\right)F_0 d^3w  = - \frac{\vth^2}{2 \nu_{ei}} \left( A_3 + A_4\right), 
\label{eqvepar}
\\
\fl q_\parallel & = \frac{1}{n_0}  \int w_\parallel  \left(\frac{\epsilon}{T_0} - \frac{5}{2}\right) \left(h_3A_3 + h_4A_4\right)F_0 d^3w  = - \frac{5 \vth^2}{4 \nu_{ei}} A_4.
\label{eqqpar}
\eea
The evolution of the perturbed electron temperature is given formally by the $\left(\epsilon / T_0 - 3/2\right)$ moment of~\eref{eqh}
\be
\left(\frac{\p}{\p{t}} + u_f \frac{\p}{\p{z}}\right)\frac{\delta T_e}{T_0} + \frac{2}{3}\frac{\p}{\p{z}}q_\parallel = i \frac{k}{B_0 l_T} \phi.
\label{eqtepert}
\ee
We now make the simplification that the equilibrium electron temperature gradient is neglected, but the ion temperature gradient is retained along with the approximation $T_e \approx T_i = T_0$. The perturbed parallel electron heat flux~\eref{eqqpar} then depends only on the perturbed electron temperature, so from~\eref{eqtepert} we see that no temperature perturbation will develop if it is initially zero. Therefore, we take here $\delta T_e = 0$.
The parallel electron flow~\eref{eqvepar} is then driven only by $A_3$ and can be replaced by the definition of the perturbed parallel current $j_\parallel = n_0 e\left(\delta V_{\parallel,i} - \delta V_{\parallel,e}\right)$. Using Amp\`{e}re's law in the Coulomb gauge, $\mu_0 j_\parallel = - \nabla^2 A_\parallel$, the expression for the electron density perturbation is obtained in terms of the five variables of interest (see~\sref{secplasmaresponse}),
\be
\fl \frac{\p}{\p{z}}\left(\frac{\delta n^t}{n_0}\right) - \frac{e}{T_0}\left(\frac{\p \phi}{\p{z}} + \frac{\p{A_\parallel}}{\p{t}} + u_f  \frac{\p A_\parallel}{\p{z}}\right) + i \frac{k}{B_0 l_n} A_\parallel = - e \frac{\eta}{\beta}\left(\frac{1}{\mu_0}\nabla_\perp^2 A_\parallel + \delta V_{\parallel,i} \right),
\label{eqqnappendix}
\ee
where the parallel resistivity is
\be
\eta = \frac{2 \nu_{ei} T_0 \beta}{n_0 e^2 \vth^2}.
\ee
Applying the normalizations defined in~\sref{secplasmaresponse} gives the quasineutrality equation~\eref{eqphipert}, with the normalized resistivity~\eref{eqresnormalised} defined as
\be
\eta_\parallel = \frac{\eta/\mu_0}{\left(c_s^2 \rho_s / l_s\right)}.
\ee

Finally the explicit parallel velocity moment of~\eref{eqh1exp} is conveniently replaced by the vorticity equation. This is readily formed by considering the Fourier representation of $h\left({\bm R}\right) = \sum_k e^{i {\bf k} \cdot {\bm R}} h_k$, multiplying the ion and electron versions of~\eref{eqh} by their respective charge and the factor $\exp \left(i {\bm k} \cdot {\bm \rho} \right)$, then summing and integrating over velocity space. The exponential factor accounts for the fact that $h$ is the distribution function of gyrocentres, and must be corrected for finite Larmor radius, before we can take advantage of quasineutrality to simplify the expression. We work from the full expression for $h$ to correctly capture terms which are formally small when determining the explicit electron response, but which cancel exactly similar ion currents. The collision terms do not contribute to this moment to leading order in the gyroradius (see the absence of collisional term in~\eref{eqnpert}) and higher order 
collisional effects are neglected, as they will appear multiplied by $\beta$, which is itself considered a small perturbation here. Neglecting nonlinear terms and denoting different species by the subscript $a = \left\{i,e\right\}$, typical contributions are evaluated as follows.
Note that with $J_{0,1}$ Bessel functions of the first kind:
\bea
\fl \frac{\p}{\p{z}} \int  e^{ik_\perp \rho} d\zeta = \frac{\p}{\p{z}} J_0\left(k_\perp \rho \right) = J_1 \left(k_\perp \rho \right) \frac{\p}{\p{z}} \left( k_\perp \rho \right) &=& J_1\left(k_\perp \rho \right) \rho \frac{\p}{\p{z}} \left(k_\perp \right) \nonumber \\ \fl & \approx & \frac{1}{2} k_y^2 \rho^2 \frac{z}{l_s^2},
\eea
where the final line follows from $J_1\left(k_\perp \rho\right) \approx k_\perp \rho / 2$ for small argument. The total contribution from the second term in~\eref{eqh} is therefore
\bea
\fl \int \sum_a e_a \left(w_\parallel + u_f\right) e^{ik_\perp \rho_a} \frac{\p}{\p z}h_a d^3w \nonumber \\
\fl = \frac{\p}{\p z} \sum_a e_a \int \left(w_\parallel + u_f \right) h_a e^{ik_\perp \rho_a} d^3w - \sum_a e_a \int \left(w_\parallel + u_f\right) h_a \frac{\p}{\p{z}} \left( e^{ik_\perp \rho_a}\right) d^3w \nonumber \\
\fl = \frac{\p}{\p z}\delta j_\parallel + u_f \sum_a \frac{n_0 e_a^2}{T_0}\frac{\p \phi}{\p z} + \frac{z}{2l_s^2} \sum_a e_a \int \left(w_\parallel + u_f\right) h_a k_\perp^2 \rho_a^2 d^3w \nonumber \\
\fl = \frac{\p}{\p z}\delta j_\parallel + 2 u_f  \frac{n_0 e^2}{T_0}\frac{\p \phi}{\p z} + \frac{n_0 e k_\perp^2 \rho_i^2}{4l_s^2}\frac{z}{1 + z^2/l_s^2}\left[\delta V_{\parallel,i} + u_f\left(\frac{\delta n_i}{n_0} + \frac{\delta T}{T_0}\right)\right],
\eea
where only the ions contribute to the final term, as the electron Larmor radius is taken to be negligible and $h \approx h^{(0)}$ given by~\eref{eqh0} is used. The remaining terms in~\eref{eqh} may be evaluated in similar fashion, taking $J_0\left(k_\perp \rho\right) \approx 1 - \left(k_\perp \rho \right)^2/4$ for small argument. Using $\mu_0 j_\parallel \approx - \nabla_\perp^2 A_\parallel$ then gives the vorticity equation
\bea
\fl \frac{\p}{\p{z}}\left(\nabla^2 A_\parallel\right) = - \frac{\p}{\p{z}}j_\parallel = k_\perp^2 \frac{n_0 T_0}{\Omega_i B^2}\left(\frac{1}{l_n} + \frac{1}{l_T}\right)\frac{\p{\phi}}{\p{y}} - \frac{k_\perp^2\rho_i^2}{2}\frac{n_0ec_s}{Bl_v}\frac{\p{A_\parallel}}{\p{y}} \nonumber \\ 
\fl + k_\perp^2 \frac{n_0e}{\Omega_i B}\left(\frac{\p}{\p{t}} + u_f \frac{\p}{\p{z}}\right)\phi + \frac{k_\perp^2\rho_i^2}{2}\frac{n_0ec_s}{l_s}\frac{z/l_s}{1+z^2/l_s^2}\left[\frac{\delta V_{\parallel,i}}{c_s} + \frac{u_f}{c_s}\left(\frac{\delta n_i}{n_0} + \frac{\delta T_i}{T_0}\right)\right].
\eea
This reduces to the form given by other authors in the appropriate limits, for example when the background flow and its shear are neglected, see~\cite{connoretal2009}. Upon applying the normalizations defined in~\sref{secplasmaresponse}, this becomes Eq.~\eref{eqapert}.


\section{Kinetic derivation of Alfv\'{e}nic instability}
\label{appendixb}

The local stability limits to parallel flow shear in the collisionless limit can be obtained from the gyrokinetic equation~\eref{hieqEdmund}, and are given here for comparison to~\eqs{eqpvgstability}{eqalfvenstability}.
To concentrate on the effect of flow shear, we neglect the background density and temperature gradients. Also, neglecting magnetic shear for this local limit, the effect of arbitrary ion Larmor radius can readily be retained and the result compared to the electrostatic case studied in~\cite{schekochihinetal2012}. We must then formally neglect the perpendicular component of the flow shear, otherwise all instabilities will automatically be convectively stabilized~\cite{schekochihinetal2012}.

Similarly to~\sref{secemldr}, we can now look for plane wave solutions of the form $\exp\left(-i\omega t + i {\bf k}\cdot {\bf x}\right)$.
Neglecting nonlinear terms, in the collisionless limit~\eref{hieqEdmund} can be rearranged directly to give the nonadiabatic piece of the distribution function for either species, denoted by subscript $a$. We allow different equilibrium ion and electron temperatures here, with ratio $\tau = T_{0e}/T_{0i}$, but for simplicity still consider singly charged ions $e_i = +e = -e_e$. Zeroth order Bessel functions appear as a result of the gyroaverage:
\be
h_a = \frac{e_aF_{0a}/T_{0a}}{k_\parallel w_\parallel - \omega}\left(\frac{w_\parallel c_sk_y}{\Omega_a l_v} -\omega \right)\left(\phi - w_\parallel A_\parallel\right)J_0\left(k_\perp\rho_a\right).
\ee
Summing the density and parallel velocity moments of the perturbed distribution over species, and requiring quasineutrality, we form the Poisson equation and parallel Amp\`{e}re's law:
\bea
\phi \sum_a \frac{n_0 e_a^2}{T_{0a}} &=& \sum_a e_a \int d^3v J_0\left(k_\perp \rho_a\right) h_a, \\
k_\perp^2 A_\parallel &=& \mu_0 \sum_a e_a  \int d^3v w_\parallel J_0\left(k_\perp \rho_a\right) h_a.
\eea
The summation must be done at the particle position, so zeroth order Bessel functions appear again, as the result of correcting the dependence of $h_a$ on the guiding centre position ${\bf R}$. Inserting the form of $h_a$ gives respectively:
\bea
\fl \left\{ 1 + \tau - \sum_a \frac{T_{0e}}{T_{0a}}\Gamma_0\left(b_a\right)\left[1-\left(1 - k_v \rho_{sa}\right)\mathcal{Z}_1\left(\overline{\omega}_a\right) \right]\right\} \phi \nonumber \\ = \sum_a \frac{T_{0e}}{T_{0a}}\Gamma_0 \left(b_a\right)\left(1 - k_v \rho_{sa}\right)\overline{\omega}_a \mathcal{Z}_1 \left(\overline{\omega}_a\right) \vtha A_\parallel, \label{eqpoisson} \\
\fl \left\{\tau b_i^2 v_A^2 - \sum_a \frac{T_{0e}}{T_{0a}} \Gamma_0 \left(b_a\right) \left[\left( 1 - k_v \rho_{sa}\right) \overline{\omega}_a^2 \mathcal{Z}_1\left(\overline{\omega}_a\right) - \frac{k_v \rho_{sa}}{2}\right]\vtha^2\right\} A_\parallel \nonumber \\ = - \sum_a \frac{T_{0e}}{T_{0a}}\Gamma_0\left(b_a\right)\left(1 - k_v \rho_{sa}\right)\overline{\omega}_a \mathcal{Z}_1\left(\overline{\omega}_a\right) \vtha \phi, \label{eqampere}
\eea
where we have defined
\bea
k_v = \frac{k_y}{k_\parallel l_v}, \qquad
\rho_{sa} = \frac{c_s}{\Omega_a}, \\
b_a = \frac{k_\perp^2 \vtha^2}{2\Omega_a^2}, \qquad
\overline{\omega}_a = \frac{\omega}{k_\parallel \vtha}.
\eea
The function $\Gamma_0 \left(x\right) = I_0\left(x\right)e^{-x}$, where $I_0$ is a modified Bessel function of the first kind. The plasma dispersion function $\mathcal{Z}_0$ and the related $\mathcal{Z}_1 = 1 + \overline{\omega}_a \mathcal{Z}_0 \left(\overline{\omega}_a\right)$ are given by~\cite{stix1992}
\be
\mathcal{Z}_n\left(\omega\right) = \frac{1}{\sqrt{\pi}}\int_{-\infty}^{\infty} \frac{\zeta^n}{\zeta - \omega}e^{-\zeta^2} d\zeta.
\ee
These two equations are a closed system at low $\beta$, and combine to give the electromagnetic dispersion relation in the collisionless limit
\bea
\fl \left\{1 + \tau - \sum_a \frac{T_{0e}}{T_{0a}}\Gamma_0\left(b_a\right)\left[1-\left(1 - k_v \rho_{sa}\right)\mathcal{Z}_1\left(\overline{\omega}_a\right) \right]\right\} \nonumber \\
\times \left\{ \tau b_i^2v_A^2 - \sum_a \frac{T_{0e}}{T_{0a}}\Gamma_0 \left(b_a\right) \left[\left( 1 - k_v \rho_{sa}\right) \overline{\omega}_a^2 \mathcal{Z}_1\left(\overline{\omega}_a\right) - \frac{k_v \rho_{sa}}{2}\vtha^2\right]\right\} \nonumber \\ 
\qquad + \frac{\omega^2}{k_\parallel^2}\left[\sum_a \frac{T_{0e}}{T_{0a}}\Gamma_0\left(b_a\right)\left(1 - k_v \rho_{sa}\right) \mathcal{Z}_1\left(\overline{\omega}_a\right)\right]^2 = 0.
\label{collisionlessdr}
\eea

In the electrostatic limit, $\beta \rightarrow 0$, the dispersion relation reduces to simply the left hand side of~\eref{eqpoisson}, equivalently the left hand factor of the first term of~\eref{collisionlessdr}, equal to zero. Further taking the massless limit for electrons, $\omega \ll k_\parallel \vthe$, $\rho_{se} \rightarrow 0$, $\Gamma_0(b_e) \rightarrow 1$ and the asymptotic form for small argument gives $\mathcal{Z}_1 \rightarrow 1$. This results in
\be
1 + \tau - \tau \Gamma_0\left(b_i\right)\left[1 - \left(1-k_v \rho_{s}\right)\left(1 + \overline{\omega}_i \mathcal{Z}_0\left(\overline{\omega}_i\right)\right)\right]=0,
\ee
which may be recognised as the dispersion relation obtained and studied previously in~\cite{schekochihinetal2012}, Eq. (16), here with $Z_i = 1$.
This can be simplified in the cold ion limit, $k_\parallel \vthi \ll \omega$ and $\Gamma_0(b_i) \rightarrow 1$. The ion acoustic wave is then weakly damped, giving the large argument form $\mathcal{Z}_1 \rightarrow - \overline{\omega}_i^{-2} + i \sqrt{\pi} \overline{\omega}_i e^{-\overline{\omega}_i^2} \equiv - \overline{\omega}_i^{-2} \left(1 - i \delta_i \right) $ and dispersion relation
\be
\omega^2 - k_\parallel^2 \frac{T_{0e}}{m_i} \left(1 - k_v \rho_s\right)\left(1-i\delta_i\right) = 0.
\ee
Compared to~\sref{secplasmaresponse}, the definition of the sound speed for unequal species temperatures is $c_s = \sqrt{\left(\gamma_e T_{0e} + \gamma_i T_{0i}\right) / m_i}$. Normalizing the frequency and length scales as in~\eref{eqnormscales}, using this more general sound speed, gives
\be
\omega^2 - k_\parallel \left(k_\parallel - k\frac{l_s}{l_v}\right)\left(1-i\delta_i\right) = 0.
\ee
In this limit, the kinetic dispersion relation may be compared to the fluid PVG dispersion relation, the first bracket in~\eref{eqwvldr}, noting that more generally $\omega_v^* = \left(\gamma_e + \gamma_i / \tau\right)^{-1} k l_s/l_v$, which reduces to $(3/8) k l_s/l_v$ for equal species temperature and $k l_s/l_v$ for cold ions.

Retaining finite $\beta$, we can also recover the Alfv\'{e}nic component of~\eref{eqwvldr}. We note that the massless electron limit first reduces~\eref{collisionlessdr} to
\bea
\fl \left\{ 1 + \tau - \tau\Gamma_0 \left(b_i\right) \left[1 - \left( 1 - k_v \rho_s\right) \mathcal{Z}_1\left(\overline{\omega}_i\right) \right]\right\} \nonumber \\
\times \left\{\tau\frac{k_\parallel^2 \vthi^2}{2\omega^2}\left[\frac{b_i}{\beta_i} - k_v\rho_s\left( 1 - \Gamma_0 \left(b_i\right)\right)\right] - 1 - \tau \Gamma_0 \left(b_i\right)\mathcal{Z}_1\left(\overline{\omega}_i\right)\left(1-k_v\rho_s\right)\right\} \nonumber \\
\qquad + \left[1 + \tau \Gamma_0 \left(b_i\right)\mathcal{Z}_1\left(\overline{\omega}_i\right)\left(1-k_v\rho_s\right)\right]^2 = 0,
\eea
where we define $\beta_i = \vthi^2/2v_A^2$. The leading terms now cancel when taking the cold ion limit, so we must retain terms to $O(b_i)$, taking $\Gamma_0(b_i) \rightarrow 1 - b_i$. Normalizing, as above, with the general sound speed, finally gives
\be
\fl \left[\omega^2 - k_\parallel \left(k_\parallel - k \frac{l_s}{l_v} \right)\left(1-i\delta_i\right)\right]\left[\omega^2 - k_\parallel \left(\frac{k_\parallel}{\beta} - \frac{\beta_i}{\beta}k \frac{l_s}{l_v} \right)\right]=0.
\ee
We recognise the kinetic equivalent to~\eref{eqwvldr}, however the effect of flow shear on the Alfv\'{e}n wave is formally small in this cold ion limit $\beta_i \rightarrow 0$, due to the assumption of adiabatic electrons.
We see that as discussed by~\cite{reynders1994} the sound and Alfv\'{e}n wave components decouple.
Retaining higher order ion Larmor radius terms would introduce the kinetic Alfv\'{e}n wave, which can couple to the ITG mode and is known to affect its stability limits. Along with the effects of Landau damping, such kinetic analysis would be required to determine the detailed frequencies and growth rates in the $M-\beta$ plot [see~\sref{secnumerics}] relevant to the core plasma.



\end{document}